\documentclass[aps,preprint,showpacs]{revtex4}

\usepackage{amsmath,amssymb}
\usepackage{bm}
\usepackage[dvips]{graphicx}

\def \G{\mathcal{G}}

\def \A{\mathcal{A}}
\def \B{\mathcal{B}}
\def \e{\varepsilon}
\def \w{\omega}
\def \k{\mathbf{k}}

\begin{document}
\title{Hot neutron matter from a Self-Consistent Green's Functions approach}
\date{\today}

\author{Arnau Rios}
\affiliation{National Superconducting Cyclotron Laboratory and Department of Physics and Astronomy, Michigan State University,
East Lansing, 48823-1 Michigan, USA}
\email[]{rios@nscl.msu.edu}

\author{Artur Polls and Isaac Vida\~na}
\affiliation{Departament d'Estructura i Constituents de la Mat\`eria and Institut
 de Ci\`{e}ncies del Cosmos, Universitat de Barcelona, Avda. Diagonal 647, E-08028 Barcelona, Spain}

\begin{abstract}
A systematic study of the microscopic and thermodynamical properties of pure neutron matter at finite temperature within the Self-Consistent Green's Function approach is performed. The model dependence of these results is analyzed by both comparing the results obtained with two different microscopic interactions,  the CD-BONN and the Argonne V18 potentials, and by analyzing the results obtained with other approaches, such as the Brueckner--Hartree--Fock approximation, the variational approach and the virial expansion. 
\end{abstract}


\maketitle

\section{Introduction}
\label{sec:intro}

The interior of neutron stars is, to a very good approximation, formed by pure neutron matter \cite{shapiro, glendenning}. At the very initial stages after their formation, these objects are very hot, with temperatures as high as $T \sim 40$ MeV \cite{prakash97}. The Equation of State (EoS) of pure neutron matter in a wide range of densities and temperatures is therefore a crucial ingredient to describe the structure and the evolution of neutron stars. The evaluation of both the neutron matter and the symmetric nuclear matter EoS starting from realistic models of the nucleon-nucleon (NN) interaction is still a major challenge in nuclear physics. The short-range and tensor components of realistic NN forces induce correlations which substantially modify the many-nucleon wave function as compared to the free Fermi gas (FFG) Slater determinant. This is particularly important for symmetric matter, where the $^3S_1$-$^3D_1$ channel plays a pivotal role. In neutron matter, Pauli effects block this tensor channel, but short-range correlations still need to be accounted for appropriately. Several theoretical approaches have been developed over the years to treat these correlations in zero temperature neutron matter: variational techniques within correlated basis functions \cite{akmal97,fantoni98,fantoni02}; Auxiliary Field \cite{gandolfi08} or Quantum Monte Carlo \cite{carlson03} calculations with simplified interactions and the popular Brueckner--Bethe--Goldstone hole-line expansion \cite{day67} in its lowest order form, the so-called Brueckner--Hartree--Fock (BHF) approximation \cite{baldo01}. At finite temperatures, fewer efforts have been focused in this direction: the well-known variational calculation of Friedman and Pandharipande \cite{friedman81} and recent similar calculations \cite{kanzawa07}, as well as BHF extensions at finite temperature \cite{cugnon87,bombaci94}. The latter approximation takes into account particle-particle correlations by solving the Bethe--Goldstone equation, which leads to the so-called $G$-matrix. Nevertheless, a minimal consistent treatment of correlations in nuclear systems requires the inclusion not only of particle-particle (pp) intermediate states, but also of the hole-hole (hh) ones. The propagation of particles and holes can be treated in the same footing by means of the Self-Consistent Green's Function (SCGF) approach \cite{muther00}. 

The SCGF approach gives direct access to the single-particle spectral function and therefore to all the single-particle properties of the system. A great progress in the application of the SCGF method to nuclear matter has been achieved in recent years, both at zero \cite{dewulf03} and finite temperatures \cite{bozek99,bozek02,frick03,frick05,rios06}. The solution of the SCGF equations is a rather demanding numerical problem due to the complete treatment of off-shell energy dependences. As a consequence, the SCGF method has been applied to few general, extensive analysis of dense nuclear systems. The studies at zero temperature have been mainly oriented to provide the appropriate theoretical support for the interpretation of $(e,e'p)$ experiments, while those at finite temperature focus on a correlated description of matter to be used in the studies of heavy ion collisions dynamics or in astrophysical environments. In particular, the effects of temperature might affect substantially different astrophysical observables. As an example, the cooling curve of a neutron star depends on the interior temperatures and the possible transition to a superfluid regime \cite{yakovlev04}. Also, the gravitational wave signature of the supernova explosion might be sensitive to the EoS and might even be able to distinguish thermal effects \cite{janka07}.

In this line, we want to study the microscopic and thermodynamical properties of hot pure neutron matter within the SCGF framework. The SCGF method, as formulated here, cannot be used below the critical temperature of the pairing transition \cite{bozek99a,dickhoff05} and therefore all our results only apply for the normal phase. Although this is not the first time that the SCGF approach is used to study pure neutron matter \cite{bozek99,dewulf03a}, it is, up to our knowledge, the first time that a systematic study of the microscopic and thermodynamical properties of pure neutron matter at finite temperature is performed within the SCGF approach. Moreover, we shall perform our calculations with two different realistic nucleon-nucleon interactions, the meson-exchange CD-BONN potential \cite{cdbonn} and the local Argonne V18 \cite{av18}. Together with the comparison to other many-body approaches, this can be used to highlight the model dependence in hot neutron matter calculations. 

Lately, the problem of neutron matter has also been growing in interest due to its connection with the experimental studies of ultracold fermionic systems \cite{carlson03,baldo08}. Dilute strongly-interacting fermionic systems with large scattering lengths (such as neutron matter, with a scattering length $a=-18$ fm to be compared to a $k_F =1.68$ fm$^{-1}$ for $\rho=0.16$ fm$^{-3}$) lie in the so-called \emph{unitary regime}. As a consequence of the lack of any characteristic energy scale, these systems show a universal behavior in their zero- and finite-temperature dynamics, with scalings that are related to the non-interacting case \cite{ho04}. We shall not treat this particular problem here, but one should mention that the SCGF method is able to tackle the unitary regime above the pairing phase transition \cite{kohler08}. When properly complemented with pairing effects \cite{bozek99a,dickhoff05}, this method should also be able to properly describe the unitary regime. 

After a brief description of the SCGF formalism in Section \ref{sec:form}, we discuss in Section \ref{sec:micro} our results for the microscopic properties of hot pure neutron matter. Section \ref{sec:macro} is devoted to the analysis of the thermodynamical properties and the comparison of our results with those obtained within other approaches. Finally, a brief summary and our main conclusions are presented in Section \ref{sec:conclu}.


\section{Self-Consistent Green's Functions method at finite temperature}
\label{sec:form}

A crucial step in the microscopic description of nuclear many-body systems is the determination of the effective in-medium nucleon-nucleon (NN) interaction. The ladder approximation to the in-medium $T$-matrix is well suited for strongly interacting low density systems \cite{fetter} and has the following structure:
\begin{align}
  \left\langle \k_1 \k_2 | T (\Omega_+) | \k_3 \k_4 \right\rangle & =  
\left\langle  \k_1 \k_2 | V | \k_3 \k_4 \right\rangle \nonumber \\
 & + \int \frac{\textrm{d}^3 k_5}{(2 \pi)^3} \frac{\textrm{d}^3 k_6}{(2 \pi)^3} 
   \left\langle \k_1 \k_2 | V | \k_5 \k_6 \right\rangle \G^0_{II}(k_5,k_6; \Omega_+)  \left\langle \k_5 \k_6 | T (\Omega_+) | \k_3 \k_4 \right\rangle \, ,
  \label{eq:lippschw}
\end{align}
where $\G^0_{II}$ is associated to the propagation of two dressed but non-interacting single-particle lines:
\begin{align}
\G^0_{II}(k,k'; \Omega_+) = \int_{-\infty}^{\infty} \frac{\textrm{d} \omega}{2 \pi} \frac{\textrm{d} \omega'}{2 \pi} 
\A(k,\w) \A(k',\w') \frac{1 - f(\w) - f(\w')}{\Omega_+ - \w -\w'} \, ,
  \label{eq:g20}
\end{align}
with $f(\w)=\left[ e^{\beta (\w - \mu)} + 1 \right]^{-1}$ the Fermi-Dirac distribution and $\A(k,\w)$ the single-particle spectral function. The notation $\Omega_\pm$ stands for $\Omega \pm i \eta$, with $\eta$ infinitesimally small. $\G^0_{II}$ can be interpreted as a Pauli blocking factor at finite temperature, analogous to the one that appears in zero temperature BHF calculations \cite{muther00}. In contrast to BHF, however, the zero temperature version of the SCGF formalism accounts for the intermediate propagation of both pp and hh states. 

The interaction of a nucleon with the remaining nucleons in the medium is described within the Green's functions formalism in terms of the self-energy \cite{kadanoff}. Its imaginary part is related to the in-medium $T$-matrix:
\begin{align}
\textrm{Im} \Sigma(k,\w) = \int \frac{\textrm{d}^3 k'}{(2 \pi)^3} \int_{-\infty}^{\infty} \frac{\textrm{d} \w'}{2 \pi} 
  \left\langle \k \k' | \textrm{Im} T (\w+\w'_+) | \k \k' \right\rangle \A(k,\w') \left[ f(\w') + b(\w+\w') \right] , 
  \label{eq:imself}
\end{align}
where a Bose-Einstein factor, $b(\Omega) = \left[ e^{-\beta(\Omega - 2\mu)} - 1 \right]^{-1}$, appears due to the symmetric treatment of pp and hh states. The real part of the self-energy is determined from its imaginary part by a dispersion relation: 
\begin{align}
\textrm{Re} \Sigma(k,\w) = \Sigma_{HF}(k)
- \mathcal{P} \int \frac{\textrm{d} \w'}{\pi} \frac{ \textrm{Im} \Sigma(k,\w'_+)}{\w-\w'} \, ,
  \label{eq:reself} 
\end{align}
except for the energy-independent Hartree-Fock contribution:
\begin{align}
\Sigma_{HF}(k) =  \int \frac{\textrm{d}^3 k'}{(2 \pi)^3} 
\left\langle \k \k' | V | \k \k' \right\rangle n(k') \, ,
\label{eq:reshf}
\end{align} 
where the momentum distribution includes the effects of correlations via $\A(k,\w)$:
\begin{align}
n(k) =  \nu \int_{-\infty}^\infty \frac{\textrm{d} \w}{2 \pi} \A(k,\w) f(\w) \, .
\label{eq:nk}
\end{align} 
$\nu=2$ accounts for the spin degeneracy of neutron matter. Finally, one can make use of Dyson's equation to close this set of equations by determining the single-particle spectral function from the real and imaginary parts of the self-energy:
\begin{align}
\A(k,\w) = \frac{-2 \textrm{Im} \Sigma(k,\w)}{ \left[\w - \frac{k^2}{2m} - \textrm{Re} \Sigma(k,\w) \right] ^2 + \left[ \textrm{Im} \Sigma(k,\w) \right]^2 } \, .
\label{eq:sf}
\end{align} 

The previous equations are derived within the grand-canonical picture, where the two external, fixed variables are the temperature, $T=\frac{1}{\beta}$, and the chemical potential, $\mu$. For dense matter studies, it is more convenient to fix the density $\rho$ and therefore we supplement the previous set of equations with the normalization condition:
\begin{align}
\rho  = \nu  \int \frac{\textrm{d}^3 k}{(2 \pi)^3}  \int_{-\infty}^\infty \frac{\textrm{d} \w}{2 \pi} \A(k,\w) f(\w,\tilde \mu) \, ,
\label{eq:rho}
\end{align} 
which determines a ``microscopic'' chemical potential, $\tilde \mu$.  In a thermodynamically consistent approximation (such as the ladder approximation),  $\tilde \mu$ should coincide with the macroscopic chemical potential, $\mu$, obtained from the bulk properties by taking the derivative of the free energy density, $F$:
\begin{align}
\mu = \left. \frac{\partial F}{\partial \rho} \right|_T \, .
\label{eq:mu}
\end{align}
Thermodynamically non-consistent many-body approximations, such as BHF, lead to $\tilde \mu \neq \mu$ \cite{baym62}. 

Equations (\ref{eq:lippschw}-\ref{eq:rho}) form a closed self-consistent set of equations in terms of the in-medium interaction, the self-energy and the single-particle spectral function that can be solved iteratively. 
The numerical details associated to the solution of these equations are rather involved and we refer the reader to Refs.~\cite{frick03,frickphd,riosphd} for further details. It is important to note that the numerical solution of the SCGF method, when available (see following paragraph), accounts for the full ladder approximation. 
 
The bosonic factor appearing in Eq.~(\ref{eq:imself}) presents a pole for $\Omega=2 \mu$, which is generally cancelled by an associated zero in $\textrm{Im} T(\Omega=2 \mu)$. However, below a certain critical temperature, $T_c$, the state with center of mass momentum $P=0$ and energy $\Omega=2 \mu$ does not cancel the bosonic factor and an instability occurs, reminiscent of the formation of Cooper pairs. This signals the onset of superfluidity, according to the so-called Thouless criterion \cite{thouless60,alm96}, and imposes a limit to the lowest temperatures we can achieve within our numerical calculations. All the results presented in the following are obtained for $T>T_c$, thus neglecting the effect of pairing correlations but guaranteeing the convergence of the approach. 

So far, we have discussed the determination of the microscopic properties of the system. The Green's function formalism can also be used to obtain the bulk properties of neutron matter. For the case of two-body interactions, one can evaluate the energy per particle by means of the Galitskii-Migdal-Koltun (GMK) sum rule \cite{migdal58,koltun74}:
\begin{align}
\frac{E}{A} = \frac{\nu}{\rho} \int \frac{\textrm{d}^3 k}{(2 \pi)^3}  \int_{-\infty}^\infty \frac{\textrm{d} \w}{2 \pi} \frac{1}{2} \left\{ \frac{k^2}{2m} + \w \right\} \A(k,\w) \, ,
\label{eq:gmk}
\end{align}
from the spectral function evaluated in the SCGF approach. To obtain the free energy, $F=E- TS$, and have a complete thermodynamical description of the system, one still needs to compute the entropy within a correlated approximation. This can be obtained by using the Luttinger-Ward (LW) formalism \cite{luttinger60b,carneiro75,rios06}. Within this approach, an expression for the grand-canonical potential in terms of dressed single-particle propagators can be obtained by means of a Legendre transformation. The entropy can then be computed from the derivative $S= - \frac {\partial \Omega}{\partial T} \mid_{\mu}$, which gives a closed expression in two terms, $S=S^{DQ}+S'$. The first one corresponds to the dynamical quasi-particle (DQ) entropy density:
\begin{align}
\frac{S_{DQ}}{A} & = \frac{\nu}{\rho} \int \frac{\textrm{d}^3 k}{(2 \pi)^3}  
\int_{-\infty}^\infty \frac{\textrm{d} \w}{2 \pi} 
 \sigma(\w) \B(k,\w) \, ,
\label{eq:sqp}
\end{align}
given by the convolution of a statistical factor, \mbox{$\sigma(\w)=-f(\w) \ln f(\w) - \left[ 1 - f(\w) \right] \ln \left[ 1-f(\w) \right]$,} and a spectral function, $\B(k,\w)$:
\begin{align}
\B(k,\w) = \left[ 1 - \frac{\partial \Sigma(k,\w)}{\partial \w} \right] 
\A(k,\w) - 2 \frac{\partial \textrm{Re} \G(k,\w)}{\partial \w} \textrm{Im} \Sigma(k,\w) \, ,
\end{align}
which can be computed from the single-particle quantities obtained in the SCGF approach. This $\B-$spectral function accounts for the effect of the dynamical (\emph{i.e.} interaction-induced) correlations that fragment the quasi-particle peak \cite{rios06}. In this paper, we will consider that the second term, $S'$, is negligible due to constraints in phase space for relatively low temperature \cite{carneiro75}. This approach leads to thermodynamical consistent results for neutron matter as well as for symmetric nuclear matter \cite{rios06,riosphd}.

In order to assess the dependence of our results on the many-body approximation employed in the description of neutron matter, we shall compare the SCGF calculations to a finite temperature generalization of the BHF method. A real finite temperature extension of the BHF approach is given by the Bloch-de Dominicis theory \cite{nicotraphd,baldo99} but, instead of using the latter, our calculations will rely on an often used simpler generalization \cite{bombaci94,rios05}. This extension can be obtained from the SCGF equations by assuming that the spectral function has no width and full strength concentrated at the BHF quasi-particle energy:
\begin{align}
\A(k,\w) = (2 \pi) \delta \left[\w - \e_{BHF}(k) \right] \, .
\end{align}
In addition, one eliminates the bosonic factor of Eq.\ (\ref{eq:lippschw}) and modifies the in-medium two-body propagator to include only intermediate particle-particle propagation: 
\begin{align}
\G^0_{II}(k,k; \Omega_+) = \frac{[1 - f(\w)][1 - f(\w')]}{\Omega_+ - \e_{BHF}(k) -\e_{BHF}(k') } \, .
  \label{eq:g20BHF}
\end{align}
The set of equations thus obtained mimics the zero temperature BHF formalism with the replacement of the step-function momentum distributions at $T=0$ by Fermi-Dirac distributions at $T \neq 0$. This guarantees that in the $T \to 0$ limit the results will coincide with BHF. One can proof that this extension coincides with the Bloch-de Dominicis results at low temperatures \cite{baldo99}.
 
Few other approaches exist that can be used to study neutron matter at finite temperatures starting from realistic NN potentials. The benchmark variational calculations of Friedman and Pandharipande (hereafter FP) \cite{friedman81} relied on a frozen correlation approximation, \emph{i.e.} using as a starting point the Jastrow-like correlation functions obtained at zero temperatures. This is of course an additional approximation, possibly only suitable for low enough temperatures and large densities, where matter can be consider degenerate. The variational approach has only recently been extended to finite temperatures to include appropriately thermal correlations \cite{mukherjee07}. Alternatively, the case of low densities and high temperatures can be studied by means of the model-independent virial expansion \cite{huang87,schwenk06}. In this approximation, the thermodynamical properties are expanded in terms of the fugacity, $z=e^{\beta \mu}$. The first term in this expansion leads to the thermodynamics of a classical free gas, while the first order correction is given in terms of a virial coefficient that can be computed from the experimental  NN interaction phase-shifts in free space. Since neutron matter is not expected to clusterize at low densities, this approximation will hold for extremely dilute and hot matter. Recently, another method has been proposed to study neutron matter at nonzero temperatures by making use of renormalized low momentum two- and three-nucleon interactions whose short-range components have been properly eliminated \cite{tolos08}. The thermal properties of neutron matter have been computed up to second order in finite temperature many-body perturbation theory, including contributions from normal and anomalous diagrams. 

\section{Microscopic properties of neutron matter}
\label{sec:micro}

In this Section we will discuss the microscopic single-particle properties of neutron matter as obtained from the SCGF approach. To address the model dependence of our calculations, we will show results using two different realistic nucleon-nucleon interactions, namely, the meson-exchange CD-BONN \cite{cdbonn} and the local Argonne V18 potentials \cite{av18}. Partial waves up to $J=8$ have been considered, with the Born approximation for $J \ge 5$ in both SCGF and BHF calculations.

\begin{figure}[t]
  \begin{center}
    \includegraphics[width=\linewidth]{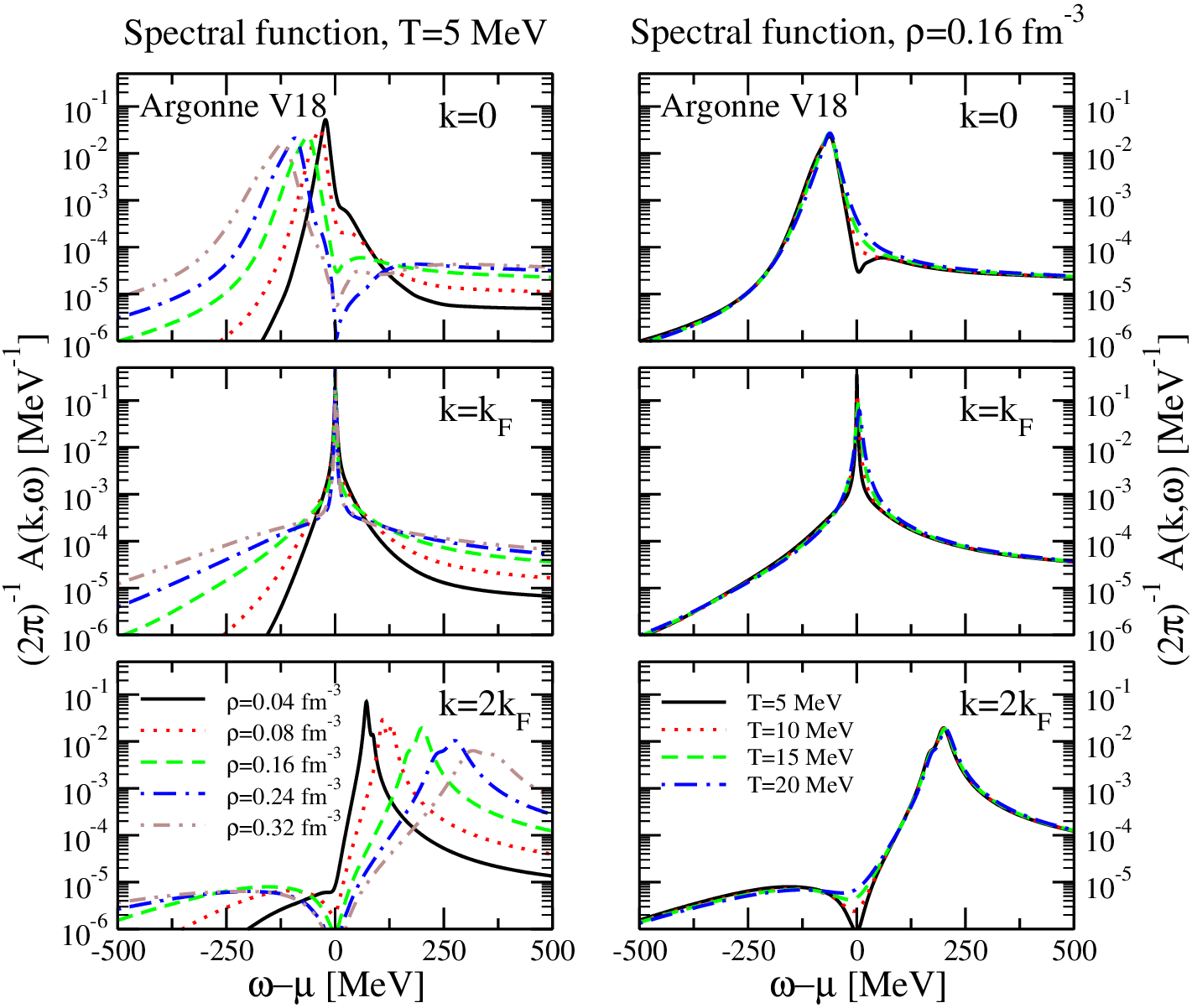}  
    \caption{(Color online) Density (left panels) and temperature (right panels) dependence of the spectral function, $\A(k,\w)$, as a function of energy for three different momenta: $k=0$ (upper), $k=k_F$ (central) and $k=2k_F$ (lower), with $k_F$ the Fermi momenta associated to each density.}
    \label{fig:asf}
  \end{center}
\end{figure}

We start by showing in Fig.~\ref{fig:asf} the density and temperature dependence of the neutron spectral function in dense neutron matter. Due to the similarity of the results for the two interactions, we will only consider the results  obtained with the Argonne V18 potential. The spectral function for densities ranging from $\rho=0.04$ fm$^{-3}$ to $\rho=0.32$ fm$^{-3}$ at a fixed temperature of $T=5$ MeV is shown in the left panels for three momenta: $k=0$ (top panel), $k=k_F$ (middle panel), and $k=2k_F$ (bottom panel). $k_F$ corresponds to the Fermi momentum associated to each density. The right panels show the results  for a fixed density, $\rho=0.16$ fm$^{-3}$, and temperatures from $T=5$ to $20$ MeV. The qualitative features of these figures are already well-known (see {\it e.g.} Ref.\ \cite{frickphd}). There is an important quasi-particle peak, which contains roughly $70-80 \%$ of the total strength for all momenta. The position of this peak changes with momenta and it is described by the self-consistent equation:
\begin{align}
  \e_{qp}(k) = \frac{\hbar^2k^2}{2m} + \textrm{Re} \, \Sigma[k,\e_{qp}(k)] \, ,
  \label{eq:qpe} 
\end{align}
which defines the quasi-particle spectrum for neutrons in the medium. With increasing density, the quasi-particle peak at zero momentum shifts to lower energies with respect to the chemical potential. It turns out that neutrons at low momenta are more bound at higher densities. The situation is the opposite for high momenta ($k \sim 2 k_F$), where the peak shifts to higher energies when density increases. At the Fermi surface, $k=k_F$, the quasi-particle peak is approximately centered around $\w \sim \mu$ and its width decreases as $\rho$ increases. At zero temperature and in the absence of pairing correlations, the spectral function would actually have a delta-like quasi-particle peak. The effect of density is particularly large in the low- and high-energy tails of the spectral function. For both large removal ($\w << \mu$) and large addition ($\w >> \mu$) energies, the strength increases with density. These off-shell components of the spectral function are populated mainly due to the action of the short-range core of the nuclear interaction and therefore it is reasonable that they increase when the mean separation between neutrons decreases. In other words, the high-energy strength of the spectral function is a good measure of the correlations induced by density effects. 

The influence of temperature in the spectral function is less pronounced. Both the position of the quasi-particle peak and the strength at low and high energies are almost unaffected by changes in temperature. The only region that is slightly modified by temperature corresponds to the range of energies $\w \sim \mu$, which is particularly sensitive to variations in phase space \cite{luttinger61}. It seems fair to say that the structure of the spectral function is mainly determined by the in-medium renormalization associated to the density, while temperature effects play a minor role. This is no longer the case close to and below $T_c$, where a relatively small decrease in temperature can lead to the appearance of superfluidity and thus to an important change in the properties of the spectral function. In particular, the onset of pairing results into a double quasi-particle peak structure close to the Fermi surface \cite{bozek99a,dickhoff05}. 

To learn more about the effect of hh propagation on the microscopic properties of neutron matter, one can compare the quasi-particle peak described by Eq.~(\ref{eq:qpe}) with the single-particle spectrum obtained within the BHF approach. The first includes both pp and hh effects, while the second only accounts for pp states. In Fig.~\ref{fig:qp} we compare the real part of the on-shell self-energy, $\textrm{Re} \Sigma[k,\e_{qp}(k)]$, for both approaches at densities $\rho=0.08$, $0.16$ and $0.24$ fm$^{-3}$ (left, central and right panel, respectively) and temperatures $T=5$ MeV (solid lines) and $T=20$ MeV (dashed lines). The results displayed in this Figure have been obtained with Argonne V18, but similar conclusions are reached with CDBONN. For all cases, the SCGF spectra are more repulsive than the BHF ones at all momenta. The effect of hh propagation in the on-shell self-energy is therefore of a repulsive nature. This effect is larger at low momenta, in accordance with the idea that the dressing induced by hh states is irrelevant for high-momentum, particle states. The repulsive effect of SCGF with respect to BHF increases with density and the differences can be as large as $25$ MeV for $k=0$ at $\rho=0.24$ fm$^{-3}$. 

The temperature behavior of the quasi-particle spectra shows some interesting features. On the one hand, the BHF single-particle spectrum becomes more repulsive with increasing temperature at all momenta. This is usually attributed to the presence of thermal Fermi-Dirac factors in the self-energy. The repulsive high relative momentum components of the interaction are not accessible at zero temperature and they only become available once the thermal distribution populates high momentum single-particle states. The overall effect is then repulsive. The same reasoning applies to particle states in the SCGF case, which also become more repulsive with increasing temperature. Hole states, on the other hand, become more attractive with increasing temperature. Presumably, this behavior can be attributed to the fact that hole states are renormalized in the SCGF, which results into a quenching of the attractive long-range components of the NN interaction in the zero temperature case. The inclusion of thermal effects leads to a somewhat weaker renormalization that increases the attractive component of the spectrum for $k<k_F$. A similar effect has been observed in extended BHF (where the repulsive contribution of holes is taken into account by the $M_2$ rearrangement term in the self-energy \cite{zuo06}) as well as in SCGF calculations of symmetric nuclear matter \cite{riosphd}.

Among the one-body properties of interest for correlated many-body systems, the momentum distribution of Eq.~(\ref{eq:nk}) is particularly sensitive to dynamical corrections. At zero temperature, for instance, the momentum distribution of the FFG is just a step-function, with complete population below $k_F$ and empty states above. In contrast, the correlated momentum distribution at $T=0$ displays a substantial depletion of hole states and a non-zero population of high momentum states. Unfortunately, the FFG at finite temperature also shows these features, since all the states become partially populated due to the thermal distribution of states. As a consequence, the correlated $n(k)$ at finite temperature will have both thermal and dynamical components. To appropriately disentangle these two components, extensive studies of the temperature and density dependence of the momentum distribution are needed.

\begin{figure}[t]
  \begin{center}
    \includegraphics[width=\linewidth]{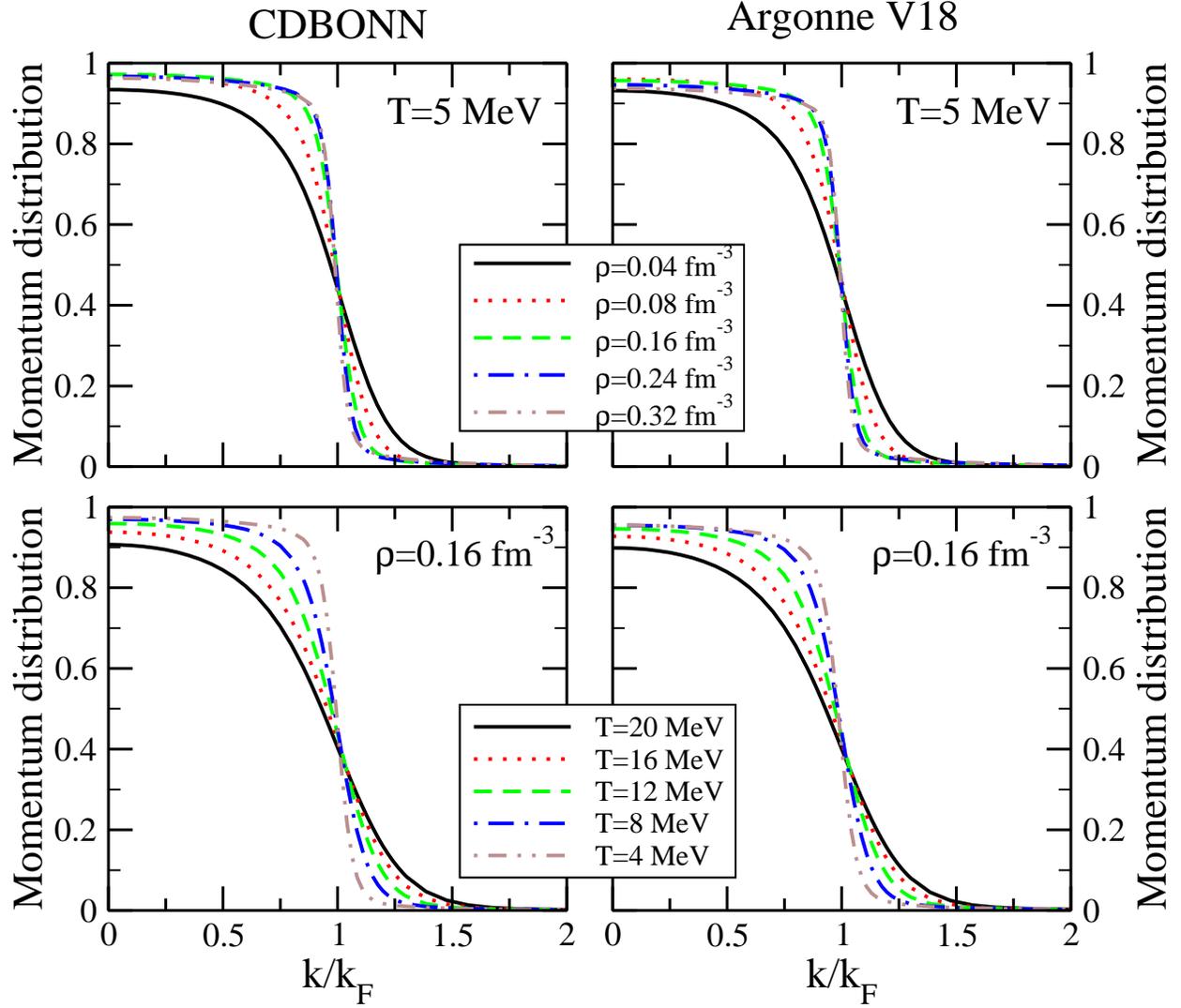}  
    \caption{(Color online) Density (top panels) and temperature (lower panels) dependence of the momentum distribution. The results are obtained with the CDBONN (left panels) and Argonne V18 (right panels) interactions.}
    \label{fig:nk}
  \end{center}
\end{figure}

This analysis is presented in Fig.\ \ref{fig:nk}, where the density (top panels) and temperature (bottom panels) dependence of the momentum distribution is shown for both CD-BONN (left panels) and Argonne V18 (right panels) potentials. Interesting analogies between the density and temperature dependence are observed: decreasing temperature has a somewhat similar effect to increasing density. This is in stark contrast to the effect of density and temperature on the spectral function, which, as already commented, are rather different. In the case of the momentum distribution, these dependences can be interpreted in terms of degeneracy arguments: for both the low temperature and the high density case, the system is reaching a degenerate limit, where thermal effects are unimportant and the depletion is essentially governed by dynamical effects. This will be the range which is interesting for understanding the influence of short-range correlations on the system. The opposite limit (low densities, high temperatures) leads to a momentum distribution which is controlled by thermal effects.

Some particular details, however, differ depending on how the degenerate limit is approached. On the one hand, fixing the density and progressively decreasing the temperature, leads to a monotonous increase (decrease) of $n(k)$ below (above) the Fermi surface. In particular, $n(k=0)$ saturates to a value different from $1$ when $T \to 0$. For $\rho=0.16$ fm$^{-3}$ at the lowest temperature available ($T=4$ MeV), one finds $n(0)=0.974$ for CDBONN and $n(0)=0.959$ for Argonne V18. The differences in the short-range components of the two interactions explain the discrepancies in $n(0)$: Argonne V18 has a harder short-range core compared to CD-BONN and thus leads, in general, to lower occupations for $k<k_F$ at high densities. On the other hand, fixing the temperature and increasing the density, one finds a different scenario, where $n(0)$ is no longer a monotonous function due to the competition of thermal and dynamical effects. 

\begin{figure}[t]
  \begin{center}
    \includegraphics[width=\linewidth]{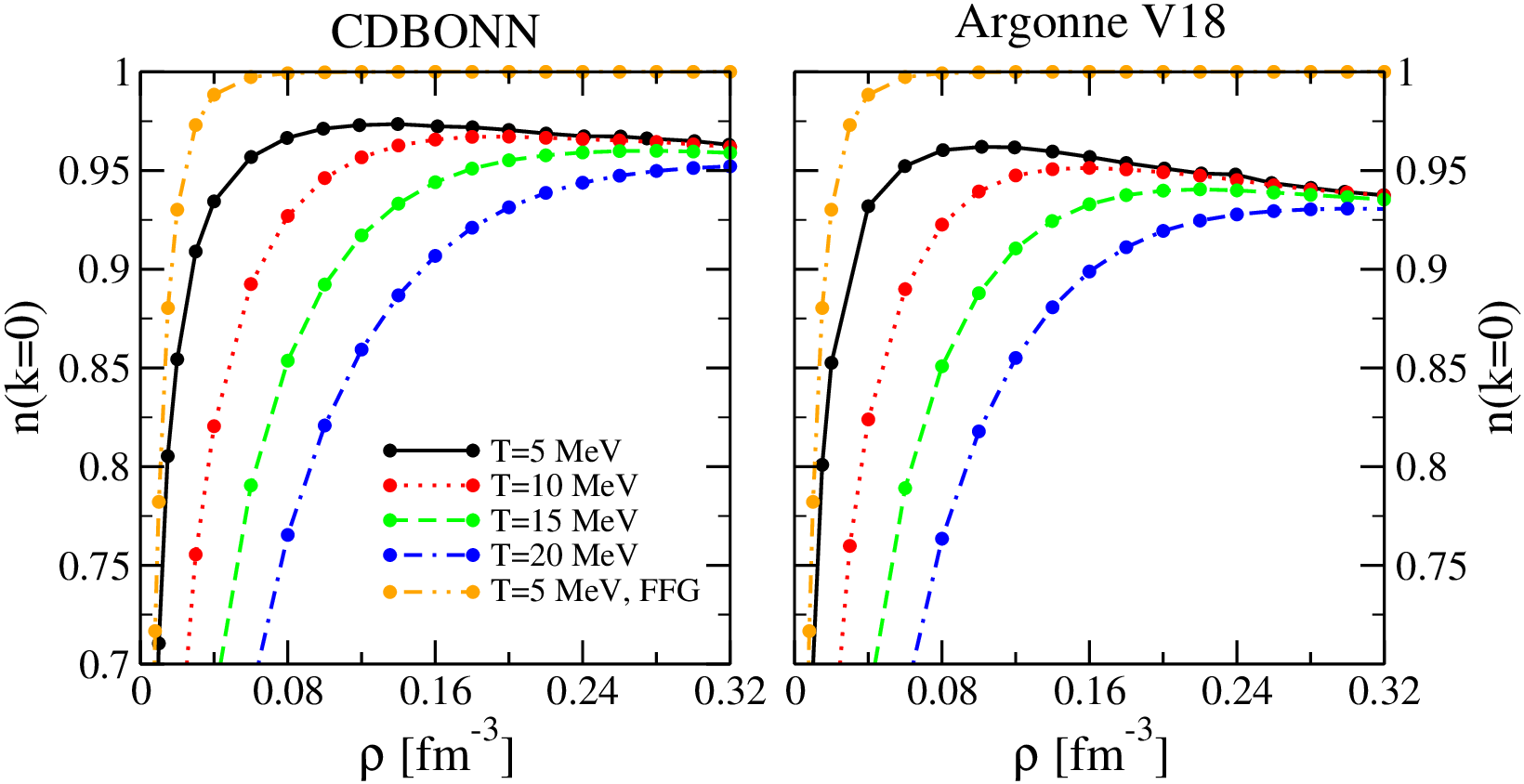}  
    \caption{(Color online) Occupation of the lowest momentum state as a function of density for four different temperatures. CDBONN (Argonne V18) results are shown on the left (right) panel. The double-dotted-dashed line corresponds to the Free Fermi Gas at $T=5$ MeV.}
    \label{fig:depl}
  \end{center}
\end{figure}

This behavior is observed in detail in Fig.\ \ref{fig:depl}, where the occupation of the lowest momentum state, $n(0)$, is shown as a function of density for several temperatures. The density dependence of $n(0)$ indeed has features which can be attributed to both thermal and dynamical effects. For all temperatures, there is a steep decrease of $n(0)$ when $\rho \to 0$. The FFG $n(0)$, shown in double-dotted dashed line, has a similar behavior, which can be explained in terms of the system approaching the classical limit ($\mu \to \infty$). In the non-interacting case, dynamical correlations are absent and therefore thermal effects are responsible for the strong decrease of $n(0)$ at low densities. The analogous behavior in the correlated $n(0)$ is basically driven by thermal correlations. The high density behavior of $n(0)$, on the other hand, is totally different from the FFG. While the latter always equals $1$ above $\rho \sim 0.08$ fm$^{-3}$ at $T=5$ MeV, the correlated $n(0)$ at this temperature tends to have values which are about $10 \%$ lower. One actually observes a decrease in $n(0)$ as density increases in this low temperature range. This dependence can be understood in terms of dynamical correlations: an increase in density results into a decrease of the mean distance between particles and, as a consequence, the importance of short-range effects is incremented at higher densities. As a consequence, the depletion increases with density, as observed. Again, this effect depends on the particular short-range structure of the NN force, which explains the differences observed between the left and right panels. Finally, let us note once again that the temperature dependence of $n(0)$ is monotonous: large temperatures lead to low $n(0)$'s at all densities. The changes induced by temperature on $n(0)$ are however density dependent and, as expected from degeneracy arguments, they are almost negligible at high densities. 

\section{Thermodynamical properties of neutron matter}
\label{sec:macro}

The SCGF approach, complemented with the Luttinger-Ward formalism, can be used to obtain the thermodynamical properties of neutron matter including the effect of correlations. In this Section we shall analyze these properties and compare the SCGF results with those of other approaches, such as the variational calculation of FP, the finite temperature extension of BHF and the virial expansion.

\begin{figure}[t]
  \begin{center}
    \includegraphics[width=\linewidth]{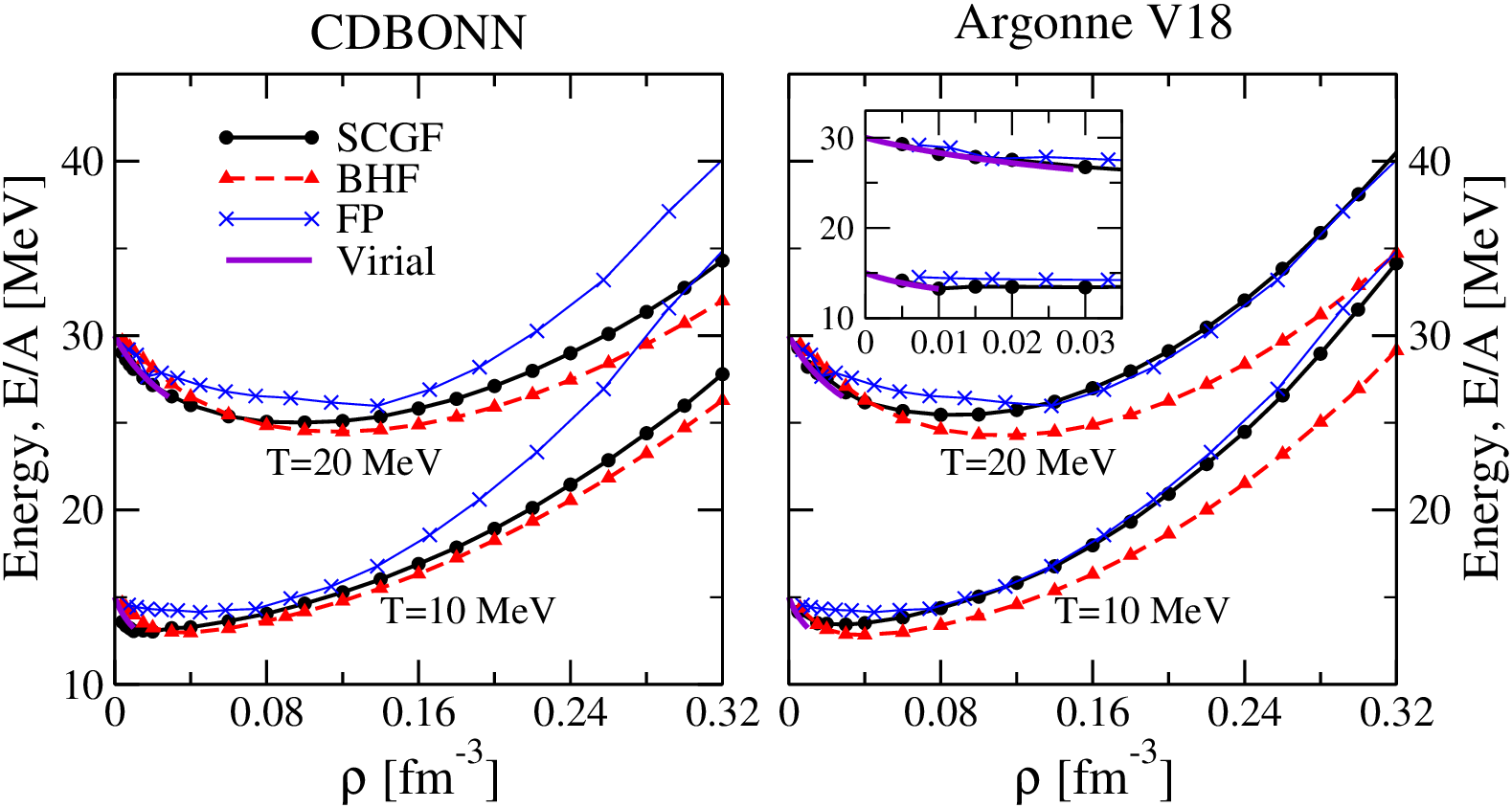}  
    \caption{(Color online) Energy per particle as a function of density for different many-body approaches: SCGF (circles), BHF (triangles), variational (crosses) and virial (solid). The results are presented at $T=10$ MeV and $T=20$ MeV. SCGF and BHF results are computed with the CDBONN (left panel) and Argonne V18 (right panel) interactions. The inset in the right panel spans the low-density regime for Argonne V18.} 
    \label{fig:ener}
  \end{center}
\end{figure}

The energy per particle, obtained from the GMK sum-rule of Eq.\ (\ref{eq:gmk}), is shown in Fig.~\ref{fig:ener} as a function of density for two temperatures, $T=10$ and $T=20$ MeV. CDBONN (Argonne V18) results are displayed in the left (right) panel. The SCGF results (circles) are compared with those obtained with the finite temperature generalization of the BHF approach (triangles), and those of the variational calculation of FP (crosses). Note that the results for the energy per particle are not quoted in the original publication and these have been obtained from the free energy and the entropy. At low densities, we also compare our results with the model-independent virial approximation for fugacities up to $z=0.5$ \cite{schwenk06}. These correspond to densities $\rho=0.0035, 0.0098, 0.0181$ and $0.0279$ fm$^{-3}$ at temperatures $T=5, 10, 15$ and $20$ MeV, respectively.

Comparing the SCGF and BHF approaches for a single NN interaction, one finds that the inclusion of hh correlations leads to a more repulsive energy per particle for almost all densities. As expected from phase space considerations, this repulsive effect is more important at higher densities. Moreover, the repulsion induced by hh propagation is more important for Argonne V18 than for CDBONN. As mentioned previously, the Argonne V18 interaction has a strong short-range core and therefore the hh renormalization on top of the pp propagation will still have an important effect. In particular, at a temperature of $T=10$ MeV, the inclusion of hh propagation leads to a $1.6$ MeV ($4.0$ MeV) increase of the energy per particle at $\rho=0.16$ fm$^{-3}$ ($0.32$ fm$^{-3}$). In contrast, the weaker short-range structure of CDBONN is already well treated with pp correlations and the inclusion of the hh component has a smaller effect, of only $0.6$ MeV ($1.5$ MeV). These results are in agreement with the zero temperature calculations of the Ghent group, which showed almost no difference between SCGF and BHF at $\rho=0.16$ fm$^{-3}$ for the Reid93 interaction \cite{dewulf03a}. However, these findings seem to disagree with those of the Krakow group \cite{bozek02a}, which suggest differences between the SCGF bulk energies and continuous choice BHF calculations of about $5$ MeV in the same conditions. Note, however, that those results were obtained with a simpler separable NN interaction and that different numerical procedures were used in the solution of the SCGF equations. The recent calculation of Ref.\ \cite{tolos08} leads to more repulsive results than ours at low densities, even when three-body effects are not considered. This is curious since, by construction, $V_{lowk}$ does not include short-range cores and thus one would have naively expected theirs EoS to be softer than that obtained by renormalizing interactions with hard cores. 

The differences in energy between the many-body approaches for a given potential that we have just discussed are a consequence of differences in the treatment of dynamical and thermal correlations. In contrast, the discrepancies within the same many-body approach for two NN interactions are a reflection of the different structure of the two potentials and, in particular, of their short-range behavior. In general, the results with Argonne V18 for both the SCGF and the BHF approaches are more repulsive than those of CDBONN. In the SCGF approach at $T=10$ MeV and $0.16$ fm$^{-3}$, for instance, $E/A=18.0$ MeV for Argonne V18, while $E/A=16.9$ MeV for CDBONN. This discrepancy increases with density and, for $\rho=0.32$ fm$^{-3}$, it becomes as large as $6.4$ MeV. In contrast, the differences in energy per particle between the two potentials for the BHF approximation are rather small. For $T=10$ MeV and at $\rho=0.16$ fm$^{-3}$, they are less than $0.5$ MeV, while at $\rho=0.32$ fm$^{-3}$ they are just about $\sim 3$ MeV. This indicates that the inclusion of hh correlations in the energy per particle increases the dependence of the results on the short-range structure of the potential. Let us also note that the discrepancy in the energy per particle due to the use of different NN potentials is somewhat larger than that associated to the use of different many-body approaches, particularly in the high density regime. 

At low densities, short-range effects are weakened and the SCGF data agrees with the virial expansion independently of the NN interaction. This is particularly well observed in the inset of Fig.~\ref{fig:ener}. This agreement provides for the first time, to the best of our knowledge, a model-independent verification of the numerics of the SCGF approach. Let us also note the relatively large differences between the FP and the SCGF results below $0.08$ fm$^{-3}$. In addition, we would like to stress that the various approximations reach the correct classical limit, $E/A \to 3T/2$ for $\rho \to 0$, but the way this limit is reached depends on the approach under consideration. In all cases, the energy per particle shows a well defined minimum. This is a consequence of the competition between thermal effects, which are dominant at low densities and tend to make the energy more repulsive, and interaction effects, which are attractive and important at intermediate densities. 

Remarkably, our SCGF results for Argonne V18 agree well with those of FP at high densities. Both calculations are based on local NN potentials, but the Urbana V14  interaction of Ref.~\cite{friedman81} includes a density-dependent quenching of the two-pion exchange contribution to account for the repulsive effect of a three-body force in a phenomenological way. Naively, one would have expected the inclusion of such a  contribution to yield more repulsive results than ours, especially at high densities. Note that, if the contribution of the three-body effects was indeed negligible, the observed agreement could be a signature of an unprecedented agreement between the variational and SCGF approaches in a wide range of temperatures and densities. To clarify this issue, it would be interesting to compare our SCGF results with finite temperature variational calculations with the Argonne V18 interaction. Alternatively, we have performed some preliminary SCGF calculations with the Urbana V14 interaction (together with the density-dependent quenching). Our preliminary calculations indicate that the energy per particle is $\sim 3$ MeV more repulsive than the SCGF energy with Argonne V18 at $\rho=0.16$ fm$^{-3}$ and $T=10$ MeV. The discrepancy increases to $\sim 10$ MeV at $0.32$ fm$^{-3}$. All in all, this seems to indicate that the agreement between the FP calculations with the Urbana V14 force and our SCGF with Argonne V18 is a coincidence, which might have been caused by some sort of cancellation between the differences induced by the two underlying interactions and those associated to the different many-body approaches. 

\begin{figure}[t]
  \begin{center}
    \includegraphics[width=\linewidth]{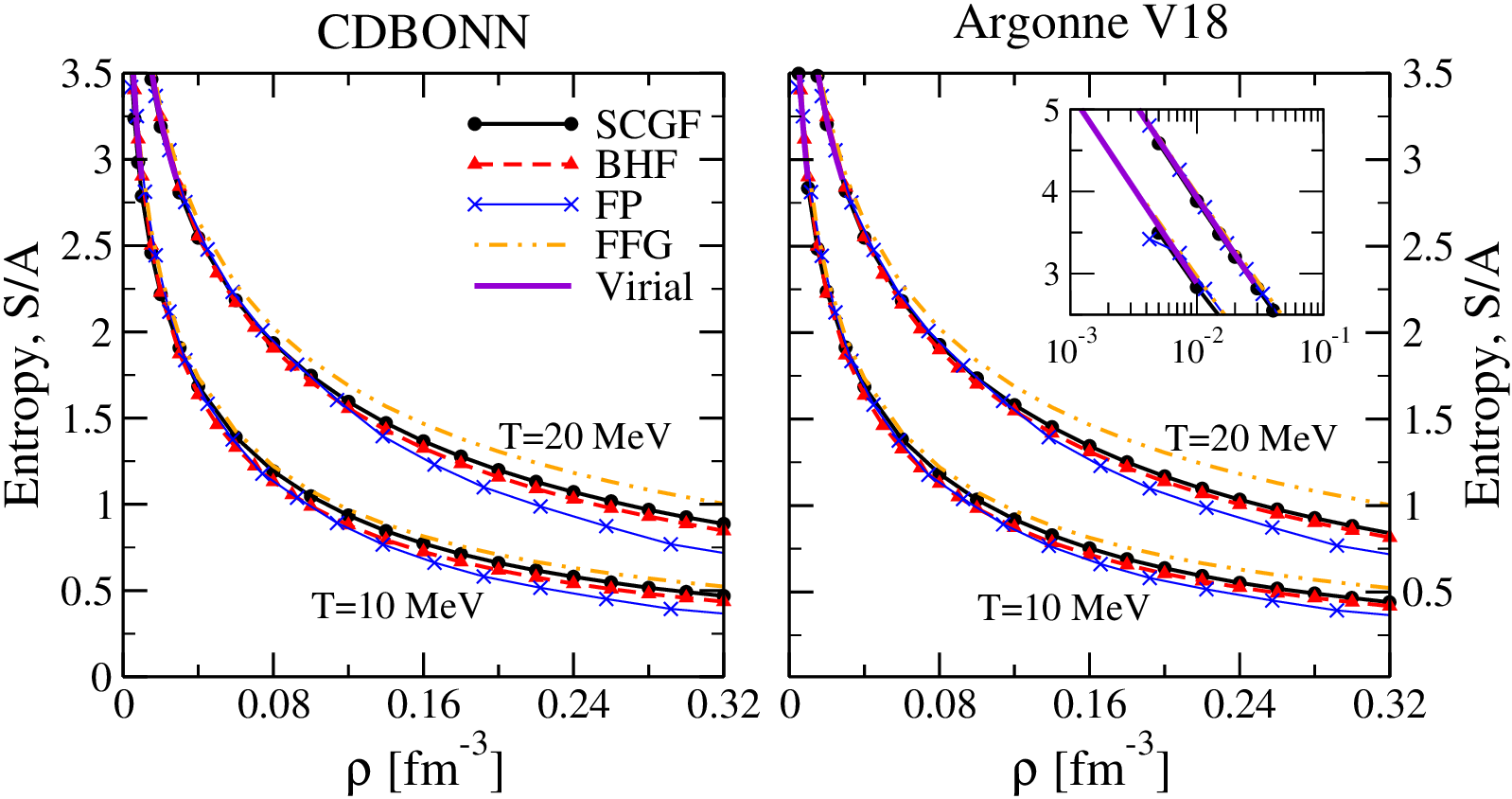}  
    \caption{(Color online) Entropy per particle as a function of density for different approximations: SCGF (circles), BHF (triangles), FP (crosses), $S_A$ (dash-dotted), FFG (double-dot dashed) and virial (solid). The results are presented at $T=10$ MeV and $T=20$ MeV. SCGF and BHF results are computed with the CDBONN (left panel) and Argonne V18 (right panel) interactions. The inset in the right panel spans the low-density regime for Argonne V18.}  
    \label{fig:entro}
  \end{center}
\end{figure}

The discrepancies are substantially smaller in the case of the entropy per particle, shown as a function of density for two temperatures, $T=10$ and $T=20$ MeV, in Fig.~\ref{fig:entro}. In particular, the changes arising from the use of different potentials (left and right panels) are smaller than those due to the use of different many-body methods.  At $T=10$ MeV, the different approximations (DQ entropy from SCGF results, BHF entropy, FP, FFG) are quite consistent with each other. The deviations above $0.16$ fm$^{-3}$ are at most of $0.15$ Boltzmann units, which would have a maximum impact on the free energy per particle of $T \times \delta S/A \sim 1.5$ MeV. At higher temperatures ($T=20$ MeV), the differences between approaches are somewhat larger, of at most $0.25$ units. 

\begin{figure}[t]
  \begin{center}
    \includegraphics[width=\linewidth]{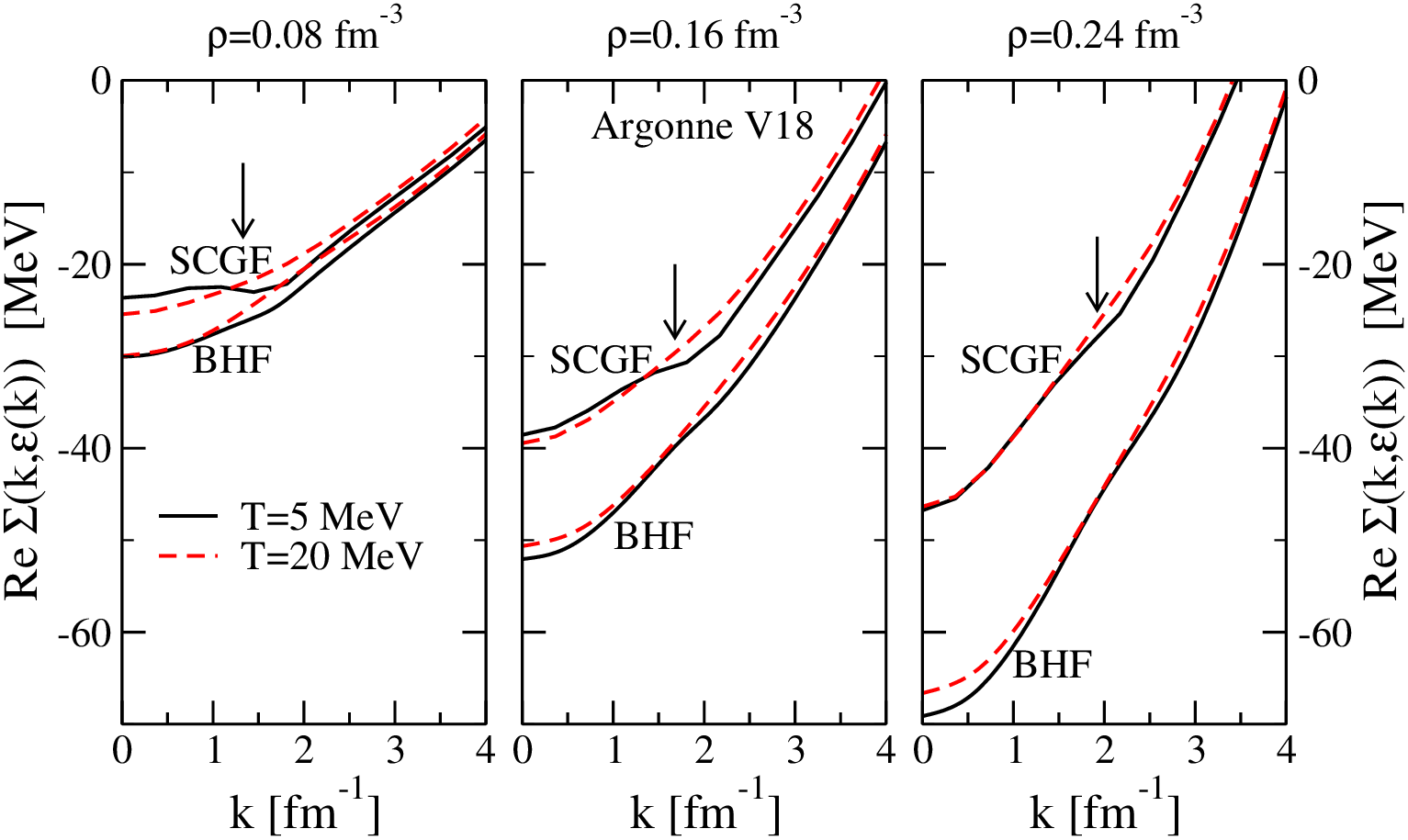}  
    \caption{(Color online) Real part of the on-shell self-energy for the SCGF and BHF approximations at $T=5$ MeV (solid lines) and $T=20$ MeV (dashed lines). The three panels correspond to densities $0.08$ fm$^{-3}$ (left), $0.16$ fm$^{-3}$ (central) and $0.24$ fm$^{-3}$ (right). The arrows show the associated Fermi momenta.}
    \label{fig:qp}
  \end{center}
\end{figure}

All in all, these results support the idea that the entropy is mostly determined by thermal correlations and rather unaffected by dynamical correlations. This is confirmed by the extremely narrow quasi-particle peak of the $\B$ spectral function. The many-body effects that fragment the quasi-particle peak, which are extremely important in the calculation of the energy, are almost negligible for the entropy \cite{rios06,riosphd}. This explains partially the good agreement between SCGF and BHF entropies. The latter are obtained by using the quasi-particle approximation to the entropy:
\begin{align}
\frac{S_{BHF}}{A} = \frac{\nu}{\rho} \int \frac{\textrm{d}^3 k}{(2 \pi)^3} \sigma \left[ \varepsilon_{BHF}(k) \right] \, .
\end{align}
Although, as observed in Fig.~\ref{fig:qp}, the quasi-particle energies of the two approaches are quite different, the change in chemical potential between BHF and SCGF shifts the entropy to values very close to those of $S_{DQ}$. The similarity between both entropies had already been observed for symmetric matter \cite{rios06}. 

\begin{figure}[t]
  \begin{center}
    \includegraphics[width=\linewidth]{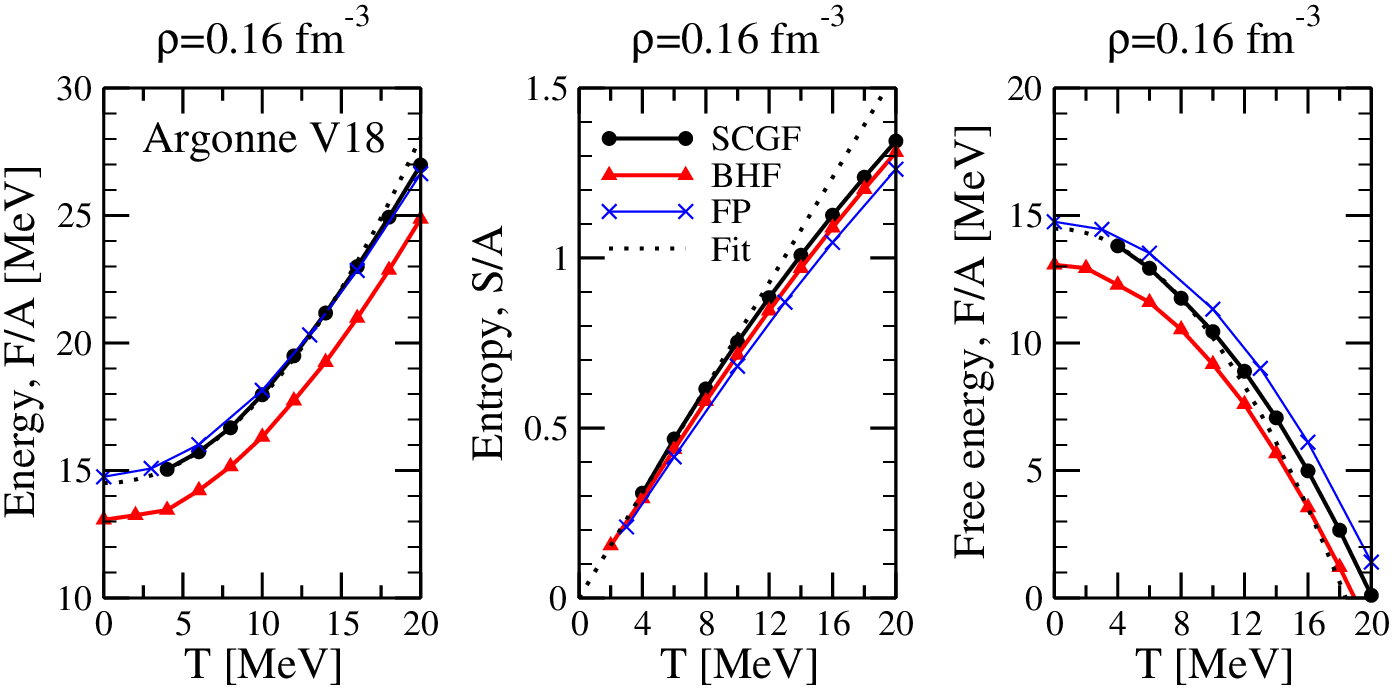}  
    \caption{(Color online) Thermodynamical properties as a function of temperature for a fixed density of $\rho=0.16$ fm$^{-3}$. The SCGF (circles), BHF (triangles) and variational (crosses) results are displayed. The dotted lines correspond to fits based on the Sommerfeld expansion.}
    \label{fig:temp}
  \end{center}
\end{figure}

Compared to the FP entropy, we find that both SCGF and BHF predict slightly larger entropies at large densities for both temperatures and interactions. A similar effect was observed in Ref.\ \cite{tolos08} and attributed to an anomalously low effective mass in variational calculations. The restriction to a quadratic spectrum in variational approaches is a limitation, especially in view of the clearly non-quadratic momentum dependences of the BHF and SCGF quasi-particle spectra (see Fig.~\ref{fig:qp}). The entropies in the latter approaches go beyond such approximation and, in the SCGF case, they even go beyond the assumption of a single quasi-particle peak. In any case, close to the degenerate limit, all the calculated entropies have a Fermi-liquid-like behavior:
\begin{align}
\frac{S}{A}= a_s T \, , 
\label{eq:sfl}
\end{align}
(see Fig.~\ref{fig:temp}), where the parameter $a_s=\frac{\pi^2 m}{\hbar^2 k_F^2}$ is given in terms of the effective mass $m^*$, calculated at the Fermi surface at zero-temperature. The discrepancy between the $S^{DQ}$ and the FP entropies at large densities (close to the degenerate limit) suggests that the effective mass in the DQ entropy is larger than that of the variational entropy. Indeed, the $m^*$ in the DQ density of states is the product of the $m^*_k$ and the $m^*_\w$ effective masses \cite{negele}. The latter is associated to the energy dependence of the self-energy and is believed to be absent in the variational approach. Since $m^*_\w$ is strongly peaked around the Fermi-surface, it leads to larger values of the total effective mass and therefore increases the DQ entropy at large densities with respect to the variational one. 

In this direction, it is important to note that, in all cases, the entropy at high densities is smaller than that predicted by the FFG. This is in accordance with the idea that, in this regime, the entropy is dominated by the effective mass, which is always smaller than $1$ and therefore leads to lower entropies. Finally, let us stress that, as expected, the interaction has little influence in the entropy near the classical regime. At low densities, all the approximations to the entropy converge to similar values and no differences are observed between the FFG and the virial entropies (see inset of Fig.~\ref{fig:entro}). 

\begin{figure}[t]
  \begin{center}
    \includegraphics[width=\linewidth]{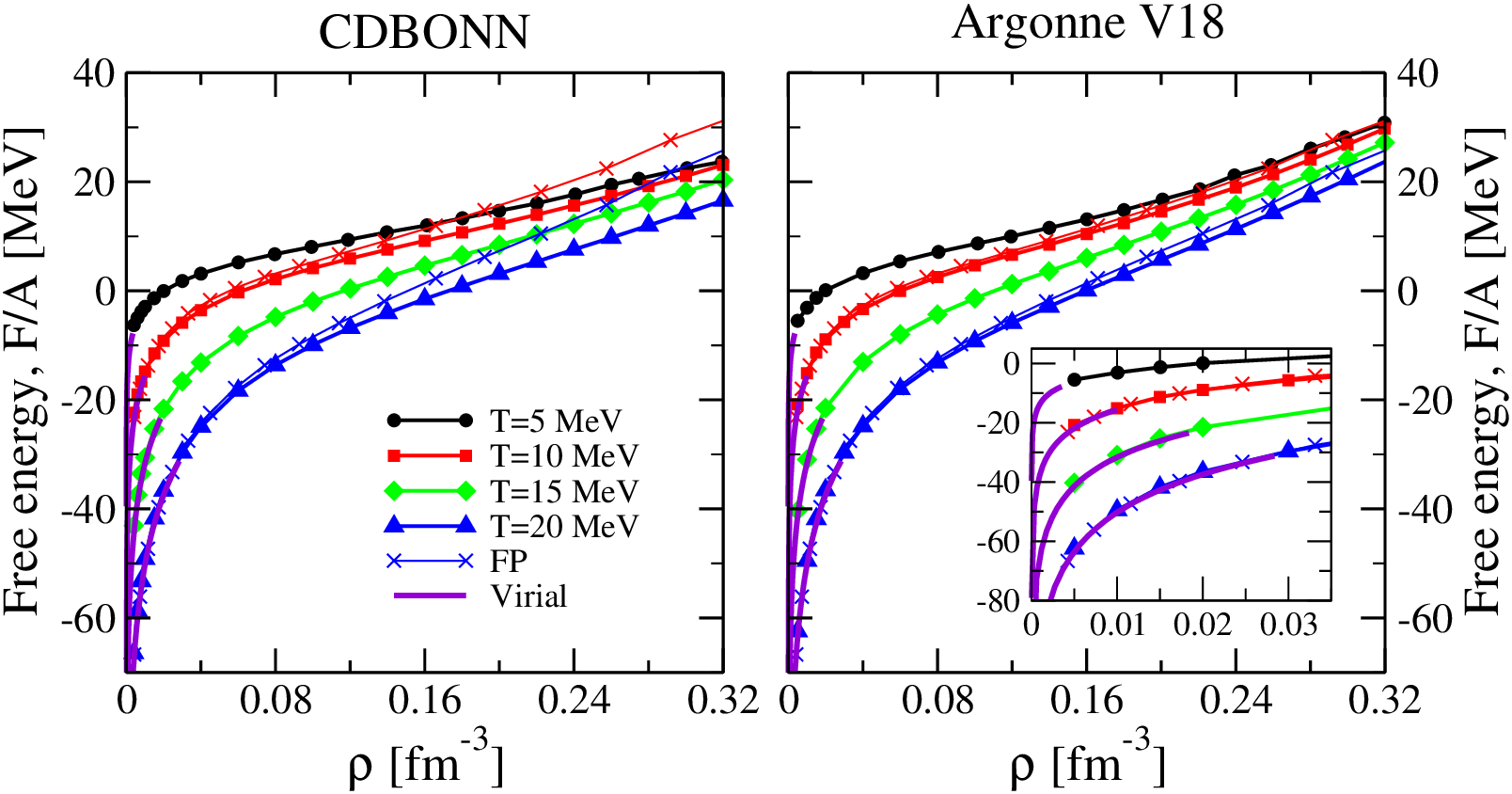}  
    \caption{(Color online) Free energy per particle as a function of density for different temperatures and different approximations: SCGF (circles), FP (crosses) and virial (solid). SCGF results are computed with the CDBONN (left panel) and Argonne V18 (right panel) interactions. The inset in the right panel spans the low-density regime for Argonne V18.}
    \label{fig:free}
  \end{center}
\end{figure}

The free-energy obtained from the GMK sum-rule complemented with the dynamical quasi-particle entropy is shown in Fig.\ \ref{fig:free} as a function of density for several temperatures. Let us first note that the calculations yield well-behaved results in a large range of densities and temperatures for both CDBONN (left panel) and Argonne V18 (right panel). In particular, the low density high-temperature regime agrees well with the virial results. This agreement is directly related to the similarity of the DQ and the virial entropies since, in this regime, the entropy overcomes the energy contribution in $F/A$. Comparing the two panels, one observes that for densities higher than $0.08$ fm$^{-3}$ the Argonne results are more repulsive than the CDBONN ones. In addition, the Argonne SCGF and the FP results are quite close to each other for all densities, with differences (mostly coming from the entropy) smaller than $3$ MeV for the highest density considered here. In general, one can say that, for low densities, $F/A$ is well determined and all the approaches agree well with each other independently of the potential. Above $\sim 0.08$ fm$^{-3}$, however, differences appear due to the sensitivity of the many-body approach to the short-range structure of each NN interaction. Let us also note that the results of Ref.~\cite{tolos08} within the two-body case are about $5-10$ MeV more repulsive than ours.

\begin{figure}[t]
  \begin{center}
    \includegraphics[width=\linewidth]{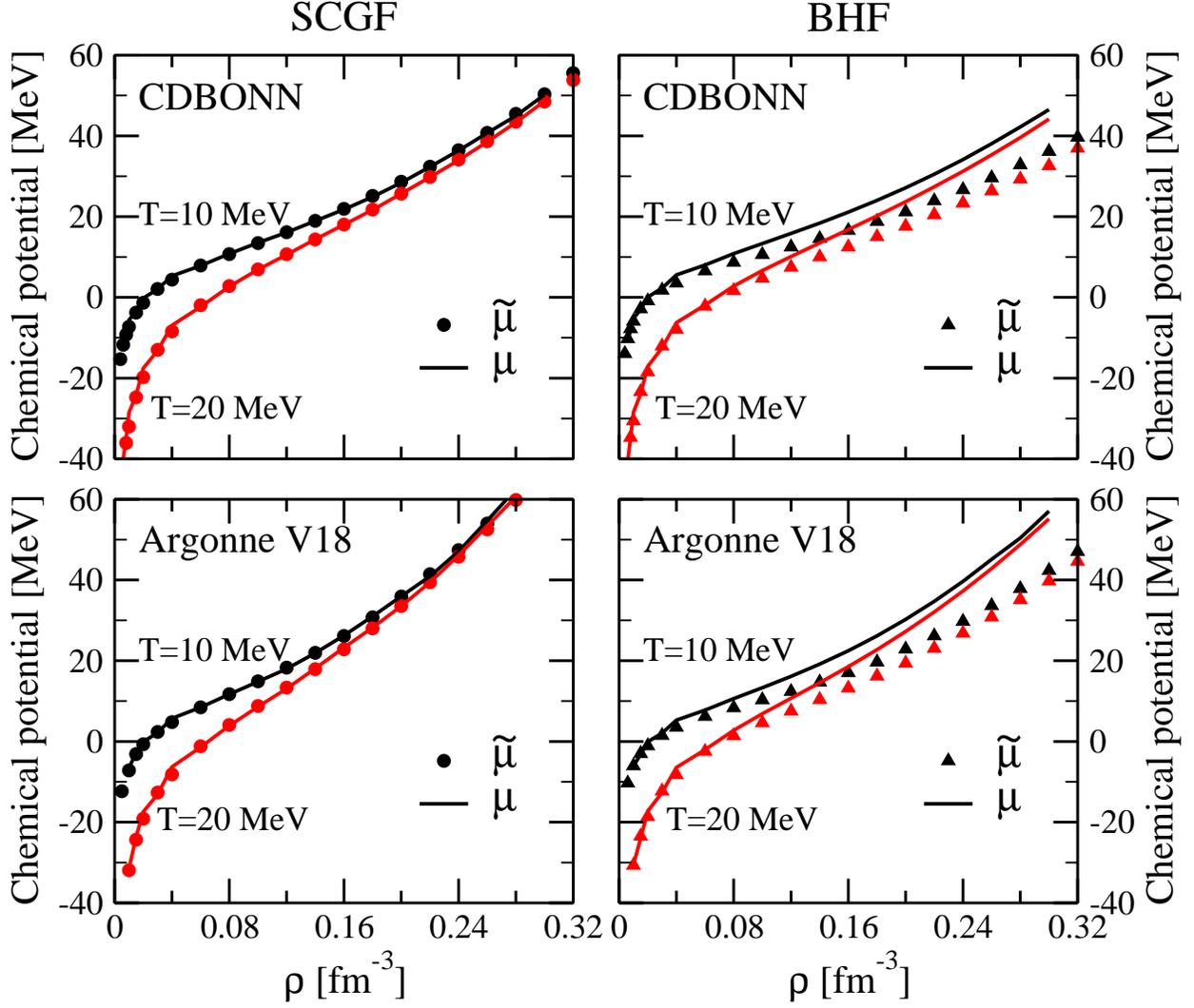}  
    \caption{(Color online) Microscopic (symbols) and macroscopic (solid lines) chemical potentials as a function of density for temperatures $T=10$ and $20$ MeV. The left (right) panels correspond to the SCGF (BHF) results, while the upper (lower) panels correspond to the CDBONN (Argonne V18) interactions.}
    \label{fig:chem}
  \end{center}
\end{figure}

Once more, we would like to stress the fact that the SCGF approach, complemented with the LW formalism, yields thermodynamically consistent results. To this end, we show in Fig.~\ref{fig:chem} the microscopic chemical potential, $\tilde \mu$, together with the macroscopic one, $\mu$, as a function of density for two temperatures, $T=10$ and $20$ MeV. Left panels correspond to SCGF results for both the CDBONN potential (upper panel) and the Argonne V18 interaction (lower panel). For the two temperatures, there is a good agreement between the microscopic chemical potential, obtained from the normalization condition of Eq.~(\ref{eq:rho}), and the macroscopic chemical potential, coming from the numerical derivative of the free energy density, Eq.~(\ref{eq:mu}). For the latter, a centered two-point formula has been used. Let us stress that this agreement confirms \emph{a posteriori} the good behavior of the dynamical quasi-particle entropy as a function of the density and also the negligible role of the $S'$ term. The approximations involved in the calculation of $S_{DQ}$ do not spoil the consistency of the ladder approximation.  Results for $\tilde \mu$ and $\mu$ within the BHF approach are presented in the right panels. Both chemical potentials agree at low ($\rho < 0.08$ fm$^{-3}$) densities for both temperatures. Above this density, however, discrepancies appear due to the increasing importance of the rearrangement contribution to the self-energy \cite{zuo06}. At $0.16$ fm$^{-3}$, the difference is of $\sim 4-5$ MeV for both potentials and temperatures, and it becomes as large as $15-20$ MeV at $0.30$ fm$^{-3}$. These differences show the lack of consistency of the BHF approach at finite temperature, even though the effect is smaller than in nuclear matter \cite{rios06}. 

\begin{figure}[t]
  \begin{center}
    \includegraphics[width=\linewidth]{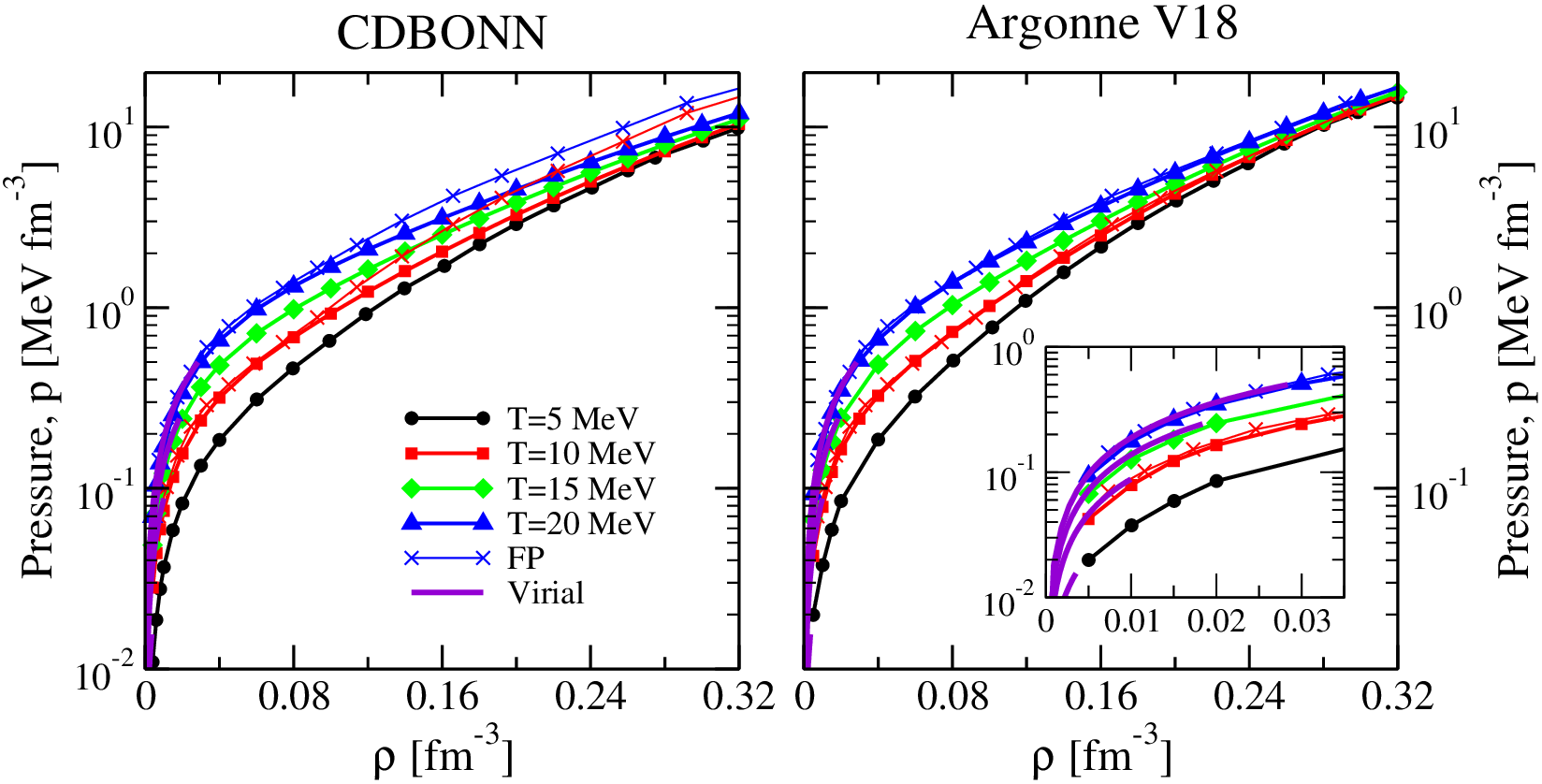}  
    \caption{(Color online) Pressure as a function of density for different temperatures and different approximations: SCGF (circles), FP (crosses) and virial (solid). SCGF results are computed with the CDBONN (left panel) and Argonne V18 (right panel) interactions. The inset in the right panel spans the low-density regime for Argonne V18.}
    \label{fig:pres}
  \end{center}
\end{figure}

The EoS of neutron matter is shown for different temperatures in Fig.\ \ref{fig:pres}. For the SCGF approach this quantity is computed from the thermodynamical relation $p=\rho \left( \tilde \mu - \frac{F}{A} \right)$, with $\tilde \mu$ the microscopic chemical potential obtained from Eq.~(\ref{eq:rho}). Note that in thermodynamically non-consistent approaches the pressure has to be computed from numerical derivatives of the free energy with respect to the density. Once again, a remarkable agreement with FP is found for the Argonne V18 results, while the CD-BONN interaction leads to a softer EoS. In the low density regime, both results agree well with the virial expansion. The effect of temperature decreases as density increases and eventually the curves for different temperatures seem to collapse to a single (density dependent) value, as expected from degeneracy arguments. This high density regime, however, will be mostly affected by the inclusion of three-body forces in the calculations. 

So far, we have plotted all our results as functions of density. To get a more accurate insight on the temperature dependence of the different thermodynamical properties of the system, we show in Fig.~\ref{fig:temp} the energy (left panel), entropy (central panel) and free energy (right panel) per particle as a function of temperature for a fixed density, $\rho=0.16$ fm$^{-3}$. The results correspond to the Argonne V18 interaction. The agreement between the SCGF and the FP energy per particle is confirmed for all temperatures (the FP results have been interpolated to this particular density). The SCGF results are about $2$ MeV more repulsive than the BHF ones and this repulsive effect is almost temperature-independent. The entropy, as expected, is well determined by all the approaches and only some small differences can be observed at large temperatures. These small differences, however, are translated into a slight disagreements between the FP and SCGF results in the free energy per particle. As observed in the right panel, the SCGF results are about $0.5$ MeV more attractive than the FP ones. The BHF results are about $1$ MeV more bound than the SCGF ones. Again, the differences between the approaches are rather temperature-independent. This suggests that the effect of dynamical correlations on the macroscopic properties are rather insensitive to thermal effects. 

It would be interesting to study the effect that sophisticated many-body calculations have on the temperature dependence of the different thermodynamical properties. This would provide a reliable test for the usually assumed quadratic (linear) temperature dependences for the energy (entropy). A detailed study of these dependences would however need of reliable extrapolations to the low-temperature regime, which in our present approach is not possible due to the presence of pairing effects. A thorough analysis of this low-temperature regime will be discussed elsewhere. At the moment, using the present data, we have parametrized the different thermodynamical quantities in terms of simple fits to study the quality of the commonly used approximations.

The energy per particle of the FFG is well fitted by a quadratic temperature dependence, $e \sim e_0 + a_e T^2$, inspired by the Sommerfeld expansion \cite{ashcroft}. This expansion is only valid for $\frac{T}{\varepsilon(k_F)} << 1$, \emph{i.e.} temperatures close to zero. In the fits to the SCGF results, however, we have to use the available data between $T=4$ and $8$ MeV. Fitting a quadratic dependence for the energy per particle of the FFG in this temperature range yields a deviation from the exact result, $a_e = \frac{\pi^2 m}{2 \hbar^2 k_F^2} = 0.0422$ MeV$^{-1}$, of only $5 \%$. Assuming that this procedure is also valid for the SCGF energies per particle, we find $a_e=0.0339$ MeV$^{-1}$. This $20 \%$ difference seems too large to be explained simply in terms of the effective mass, which in this regime is $m^* \sim 0.9 m$. The naive replacement $a_e=\frac{\pi^2 m^*}{2 \hbar^2 k_F^2}$, for instance, does not agree with the previous value. The more accurate prediction $\frac{\pi^2 m^*}{2 \hbar^2 k_F^2} \frac{m^*+m}{2m}$ \cite{grange87}, although closer, is also somewhat too large to coincide with the fit to SCGF data. Alternatively, one could have tried to obtain an analytic expression for $a_e$ from the low temperature expansion of the GMK sum rule formula, but this is difficult due to the non-trivial temperature dependence of $\A(k,\w)$. Let us also stress that the quadratic thermal dependence of the energy per particle comes essentially from the kinetic energy term. The potential energy is rather temperature independent and decreases by less than $2$ MeV when going from $T=20$ to $4$ MeV. 

According to Fermi liquid theory, the behavior of the entropy at low temperatures should be linear with $T$. For the FFG, a fit of Eq.~(\ref{eq:sfl}) in the $T=4$ to $8$ MeV regime gives a very accurate value of $a_s=2 a_e=0.0844$ MeV$^{-1}$. A similar one-parameter fit to the SCGF yields a slope, $a_s \sim 0.0772$ MeV$^{-1}$, in agreement with the Fermi liquid prediction for an effective mass, $m^* \sim 0.92 m$. This coincides with the value that we obtain for the effective mass at $k=k_F$ at low temperatures. The entropy however shows a clear deviation from this linear behavior above $12$ MeV. In addition, the FFG prediction $a_s = 2 a_e$ is partially violated. Finally, a quadratic fit to the SCGF data for the free energy per particle, $f=f_0 + a_f T^2$, leads to $a_f=-0.0428$ MeV $^{-1}$. This is a somewhat low value, rather close to the FFG prediction. In contrast, the FFG relation $a_e=-a_f$ is not well fulfilled. Moreover, the non-linear behavior of the entropy for $T>12$ MeV leads to a non-quadratic behavior of $F/A$ above this temperature. A consistency check of these low temperature fits is the relation $a_e - a_s \sim a_f$ as well as the fact that the zero-temperature extrapolation of $E/A$, $e_0 = 14.51$, and of $F/A$, $f_0 = 14.49$ MeV, do coincide. Note, however, that the accuracy of the fits depend on the exact position and the number of points considered at low temperatures and these are the most sensitive to numerical uncertainties within our approach. Finally, we would like to stress that the convergence of our results down to $T=4$ MeV implies that ${T_c < 4}$ MeV.


\section{Conclusions}
\label{sec:conclu}

We have presented the first systematic study of hot neutron matter within the Self-Consistent Green's Function formalism in the ladder approximation for two realistic NN interactions, the CDBONN and the Argonne V18 potentials. The calculations cover a wide range of densities and temperatures and show the adequateness of this method to account for correlations in the microscopic properties of dense, hot hadronic matter. The effect of short range correlations in the thermodynamical properties is correctly described by the Luttinger-Ward formalism.  

At the microscopic level, short-range effects are particularly important on the spectral functions and are manifested in its low and high energy tails. Our results indicate that both the location of the quasi-particle peak and the amount of strength in the energy tails change substantially with density. On the contrary, thermal effects are very small and only affect the region around the chemical potential. The momentum dependence of the real part of the on-shell self-energy in the SCGF approach has been compared to the single-particle spectrum in BHF-like descriptions. The propagation of hh pairs in the intermediate states has a repulsive effect with respect to BHF results. This difference grows with density and is larger for momenta below the Fermi momentum, with maximum differences of $\sim 25$ MeV in the range of densities explored here. In the hole-momentum region, in addition, BHF and SCGF show different thermal behaviors, with the latter becoming more attractive as temperature increases. 

A careful study of the momentum distribution of the system has also been performed and the important interplay between thermal and dynamical correlations has been highlighted. These effects are well exemplified by $n(0)$, which is customarily used as a measure of correlation effects. For a given temperature and decreasing density, the system approaches the classical limit and the depletion of the momentum distribution increases. For larger densities, closer to the degenerate limit, dynamical correlations play a more important role and $n(0)$ decreases with increasing density. In general, correlation effects, as measured by the depletion, are larger for Argonne V18 than for CDBONN. 

In general, the SCGF energy per particle is more repulsive than the BHF one, independently of the interaction. The magnitude of these differences is governed by the density and by the particular structure of the NN interaction, and it is at most of $5$ MeV for the range of densities explored (up to $0.32$ fm$^{-3}$). The sensitivity to the NN interaction within the SCGF approach appears to be larger than this, with differences of up to $6$ MeV in the same range. In contrast, BHF results are relatively potential-independent. In any case, in the low density regime the energies for both approaches compare very well with the virial expansion.  In addition, there is a very good agreement between the SCGF results for Argonne V18 and those of FP for Urbana V14 above a density of $0.08$ fm$^{-3}$. This is possibly due to a cancellation between the potential and the many-body dependence of the energy per particle in this regime. 

The entropy has been computed within the dynamical quasi-particle approximation, which takes into account the effects of correlations in the width of the quasi-particle peak. The discrepancies between different approximations to the entropy are rather small, thus revealing that the entropy is not affected by correlations. In general, all the approaches lead to somewhat lower values than those predicted by the FFG. The free energy for Argonne V18 and CDBONN shows substantial differences at large densities due to the different structures of the potential. The free energy obtained from the GMK energy and the DQ entropy leads to a thermodynamical consistent result, with a good agreement between the microscopic and the macroscopic chemical potentials. The differences for the BHF approach can be as large as $20$ MeV, although in general they are less important than for the nuclear matter case. The EoS, which has been computed in a wide range of densities and temperatures, also shows a similar potential dependence. In the low density regime, however, all the thermodynamical quantities show a very good agreement with the virial expansion. The stability of our results in this regime shows the robustness of the numerical techniques involved in the calculations. 

The temperature dependence of the energy, the entropy and the free energy has also been explored. A quadratic dependence is compatible for the energy per particle at $\rho=0.16$ fm$^{-3}$ for the SCGF and BHF approaches.  The entropy is only proportional to the temperature below $T \sim 10$ MeV, which in turn translates into a non-quadratic temperature dependence of the free energy above this temperature. Moreover, the differences in energy and free energy between BHF and SCGF remain constant with temperature, indicating that the effect of temperature on dynamical correlations is rather small. In addition, the convergence of the results down to $T=4$ MeV indicate that no superfluidity appears above this temperature. In conclusion, the calculations performed show the potential of the SCGF method to describe accurately the properties of dense and hot matter. The inclusion of pairing effects and three-body forces within this formalism will improve the predictions for the microscopic and the bulk properties and will provide a very complete description of neutron star matter. 

\section{Acknowledgments}

This work was partially supported by the NSF under Grant No. PHY-0555893, the MEC (Spain) and FEDER under Grant No. FIS2005-03142, and by the Generalitat de Catalunya (Spain) under Grant No. 2005SGR-00343.

\bibliographystyle{apsrev}
\bibliography{main}

\begin{thebibliography}{55}
\expandafter\ifx\csname natexlab\endcsname\relax\def\natexlab#1{#1}\fi
\expandafter\ifx\csname bibnamefont\endcsname\relax
  \def\bibnamefont#1{#1}\fi
\expandafter\ifx\csname bibfnamefont\endcsname\relax
  \def\bibfnamefont#1{#1}\fi
\expandafter\ifx\csname citenamefont\endcsname\relax
  \def\citenamefont#1{#1}\fi
\expandafter\ifx\csname url\endcsname\relax
  \def\url#1{\texttt{#1}}\fi
\expandafter\ifx\csname urlprefix\endcsname\relax\def\urlprefix{URL }\fi
\providecommand{\bibinfo}[2]{#2}
\providecommand{\eprint}[2][]{\url{#2}}

\bibitem[{\citenamefont{Shapiro and Teukolsky}(1983)}]{shapiro}
\bibinfo{author}{\bibfnamefont{S.~L.} \bibnamefont{Shapiro}} \bibnamefont{and}
  \bibinfo{author}{\bibfnamefont{S.~A.} \bibnamefont{Teukolsky}},
  \emph{\bibinfo{title}{Black holes, white dwarfs and neutron stars: The
  physics of compact objects}} (\bibinfo{publisher}{John Wiley and Sons},
  \bibinfo{address}{N.Y.}, \bibinfo{year}{1983}).

\bibitem[{\citenamefont{Glendenning}(2000)}]{glendenning}
\bibinfo{author}{\bibfnamefont{N.~K.} \bibnamefont{Glendenning}},
  \emph{\bibinfo{title}{Compact Stars: Nuclear Physics, Particle Physics and
  General Relativity}} (\bibinfo{publisher}{Springer}, \bibinfo{year}{2000}),
  \bibinfo{edition}{2nd} ed.

\bibitem[{\citenamefont{Prakash et~al.}(1997)}]{prakash97}
\bibinfo{author}{\bibfnamefont{M.}~\bibnamefont{Prakash}} \bibnamefont{et~al.},
  \bibinfo{journal}{Phys. Reps.} \textbf{\bibinfo{volume}{280}},
  \bibinfo{pages}{1} (\bibinfo{year}{1997}).

\bibitem[{\citenamefont{Akmal and Pandharipande}(1997)}]{akmal97}
\bibinfo{author}{\bibfnamefont{A.}~\bibnamefont{Akmal}} \bibnamefont{and}
  \bibinfo{author}{\bibfnamefont{V.~R.} \bibnamefont{Pandharipande}},
  \bibinfo{journal}{Phys. Rev. C} \textbf{\bibinfo{volume}{56}},
  \bibinfo{pages}{2261} (\bibinfo{year}{1997}).

\bibitem[{\citenamefont{Fantoni and Fabrocini}(1998)}]{fantoni98}
\bibinfo{author}{\bibfnamefont{S.}~\bibnamefont{Fantoni}} \bibnamefont{and}
  \bibinfo{author}{\bibfnamefont{A.}~\bibnamefont{Fabrocini}}, in
  \emph{\bibinfo{booktitle}{Microscopic Quantum Many-Body Theories and Their
  Applications}}, edited by
  \bibinfo{editor}{\bibfnamefont{J.}~\bibnamefont{Navarro}} \bibnamefont{and}
  \bibinfo{editor}{\bibfnamefont{A.}~\bibnamefont{Polls}}
  (\bibinfo{publisher}{Springer}, \bibinfo{address}{New York},
  \bibinfo{year}{1998}).

\bibitem[{\citenamefont{Fantoni}(2002)}]{fantoni02}
\bibinfo{author}{\bibfnamefont{S.}~\bibnamefont{Fantoni}}, in
  \emph{\bibinfo{booktitle}{Introduction to Modern Methods of Quantum Many-Body
  Theory and Their Applications}}, edited by
  \bibinfo{editor}{\bibfnamefont{A.}~\bibnamefont{Fabrocini}},
  \bibinfo{editor}{\bibfnamefont{S.}~\bibnamefont{Fantoni}}, \bibnamefont{and}
  \bibinfo{editor}{\bibfnamefont{E.}~\bibnamefont{Krotscheck}}
  (\bibinfo{publisher}{Singapore}, \bibinfo{address}{World Scientific},
  \bibinfo{year}{2002}), vol.~\bibinfo{volume}{7} of
  \emph{\bibinfo{series}{Series on Advances in Quantum Many-Body Theory}}.

\bibitem[{\citenamefont{Gandolfi et~al.}()\citenamefont{Gandolfi, Illarionov,
  Fantoni, Pederiva, and Schmidt}}]{gandolfi08}
\bibinfo{author}{\bibfnamefont{S.}~\bibnamefont{Gandolfi}},
  \bibinfo{author}{\bibfnamefont{A.~Y.} \bibnamefont{Illarionov}},
  \bibinfo{author}{\bibfnamefont{S.}~\bibnamefont{Fantoni}},
  \bibinfo{author}{\bibfnamefont{F.}~\bibnamefont{Pederiva}}, \bibnamefont{and}
  \bibinfo{author}{\bibfnamefont{K.~E.} \bibnamefont{Schmidt}},
  \bibinfo{note}{arxiv:0805.2513}.

\bibitem[{\citenamefont{Carlson et~al.}(2003)\citenamefont{Carlson, Morales,
  Pandharipande, and Ravenhall}}]{carlson03}
\bibinfo{author}{\bibfnamefont{J.}~\bibnamefont{Carlson}},
  \bibinfo{author}{\bibfnamefont{J.}~\bibnamefont{Morales}},
  \bibinfo{author}{\bibfnamefont{V.~R.} \bibnamefont{Pandharipande}},
  \bibnamefont{and} \bibinfo{author}{\bibfnamefont{D.~G.}
  \bibnamefont{Ravenhall}}, \bibinfo{journal}{Phys. Rev. C}
  \textbf{\bibinfo{volume}{68}}, \bibinfo{pages}{025802}
  (\bibinfo{year}{2003}).

\bibitem[{\citenamefont{Day}(1967)}]{day67}
\bibinfo{author}{\bibfnamefont{B.~D.} \bibnamefont{Day}},
  \bibinfo{journal}{Rev. Mod. Phys.} \textbf{\bibinfo{volume}{39}},
  \bibinfo{pages}{719} (\bibinfo{year}{1967}).

\bibitem[{\citenamefont{Baldo and Burgio}(2001)}]{baldo01}
\bibinfo{author}{\bibfnamefont{M.}~\bibnamefont{Baldo}} \bibnamefont{and}
  \bibinfo{author}{\bibfnamefont{G.~F.} \bibnamefont{Burgio}}, in
  \emph{\bibinfo{booktitle}{Microscopic Theory of Nuclear Equation of State and
  Neutron Star Structure}}, edited by
  \bibinfo{editor}{\bibfnamefont{D.}~\bibnamefont{Blaschke}},
  \bibinfo{editor}{\bibfnamefont{N.~K.} \bibnamefont{Glendenning}},
  \bibnamefont{and} \bibinfo{editor}{\bibfnamefont{A.}~\bibnamefont{Sedrakian}}
  (\bibinfo{publisher}{Heidelberg}, \bibinfo{address}{Springer},
  \bibinfo{year}{2001}), vol. \bibinfo{volume}{578} of
  \emph{\bibinfo{series}{Lecture Notes in Physics}}.

\bibitem[{\citenamefont{Friedman and Pandharipande}(1981)}]{friedman81}
\bibinfo{author}{\bibfnamefont{B.}~\bibnamefont{Friedman}} \bibnamefont{and}
  \bibinfo{author}{\bibfnamefont{V.~R.} \bibnamefont{Pandharipande}},
  \bibinfo{journal}{Nucl. Phys. A} \textbf{\bibinfo{volume}{361}},
  \bibinfo{pages}{502} (\bibinfo{year}{1981}).

\bibitem[{\citenamefont{Kanzawa et~al.}(2007)\citenamefont{Kanzawa, Oyamatsu,
  Sumiyoshi, and Takanoe}}]{kanzawa07}
\bibinfo{author}{\bibfnamefont{H.}~\bibnamefont{Kanzawa}},
  \bibinfo{author}{\bibfnamefont{K.}~\bibnamefont{Oyamatsu}},
  \bibinfo{author}{\bibfnamefont{K.}~\bibnamefont{Sumiyoshi}},
  \bibnamefont{and} \bibinfo{author}{\bibfnamefont{M.}~\bibnamefont{Takanoe}},
  \bibinfo{journal}{Nucl. Phys. A} \textbf{\bibinfo{volume}{791}},
  \bibinfo{pages}{232} (\bibinfo{year}{2007}).

\bibitem[{\citenamefont{Cugnon et~al.}(1987)\citenamefont{Cugnon, Deneye, and
  Lejeune}}]{cugnon87}
\bibinfo{author}{\bibfnamefont{J.}~\bibnamefont{Cugnon}},
  \bibinfo{author}{\bibfnamefont{P.}~\bibnamefont{Deneye}}, \bibnamefont{and}
  \bibinfo{author}{\bibfnamefont{A.}~\bibnamefont{Lejeune}},
  \bibinfo{journal}{Z. Phys. A} \textbf{\bibinfo{volume}{328}},
  \bibinfo{pages}{409} (\bibinfo{year}{1987}).

\bibitem[{\citenamefont{Bombaci et~al.}(1994)\citenamefont{Bombaci, Kuo, and
  Lombardo}}]{bombaci94}
\bibinfo{author}{\bibfnamefont{I.}~\bibnamefont{Bombaci}},
  \bibinfo{author}{\bibfnamefont{T.~T.~S.} \bibnamefont{Kuo}},
  \bibnamefont{and} \bibinfo{author}{\bibfnamefont{U.}~\bibnamefont{Lombardo}},
  \bibinfo{journal}{Phys. Reps.} \textbf{\bibinfo{volume}{242}},
  \bibinfo{pages}{165} (\bibinfo{year}{1994}).

\bibitem[{\citenamefont{M{\"u}ther and Polls}(2000)}]{muther00}
\bibinfo{author}{\bibfnamefont{H.}~\bibnamefont{M{\"u}ther}} \bibnamefont{and}
  \bibinfo{author}{\bibfnamefont{A.}~\bibnamefont{Polls}},
  \bibinfo{journal}{Prog. Part. Nucl. Phys.} \textbf{\bibinfo{volume}{45}},
  \bibinfo{pages}{243} (\bibinfo{year}{2000}).

\bibitem[{\citenamefont{Dewulf et~al.}(2003)\citenamefont{Dewulf, Dickhoff,
  Van~Neck, Stoddard, and Waroquier}}]{dewulf03}
\bibinfo{author}{\bibfnamefont{Y.}~\bibnamefont{Dewulf}},
  \bibinfo{author}{\bibfnamefont{W.~H.} \bibnamefont{Dickhoff}},
  \bibinfo{author}{\bibfnamefont{D.}~\bibnamefont{Van~Neck}},
  \bibinfo{author}{\bibfnamefont{E.~R.} \bibnamefont{Stoddard}},
  \bibnamefont{and}
  \bibinfo{author}{\bibfnamefont{M.}~\bibnamefont{Waroquier}},
  \bibinfo{journal}{Phys. Rev. Lett.} \textbf{\bibinfo{volume}{90}},
  \bibinfo{pages}{152501} (\bibinfo{year}{2003}).

\bibitem[{\citenamefont{Bo{\.z}ek}(1999{\natexlab{a}})}]{bozek99}
\bibinfo{author}{\bibfnamefont{P.}~\bibnamefont{Bo{\.z}ek}},
  \bibinfo{journal}{Phys. Rev. C} \textbf{\bibinfo{volume}{59}},
  \bibinfo{pages}{2619} (\bibinfo{year}{1999}{\natexlab{a}}).

\bibitem[{\citenamefont{Bo{\.z}ek}(2002)}]{bozek02}
\bibinfo{author}{\bibfnamefont{P.}~\bibnamefont{Bo{\.z}ek}},
  \bibinfo{journal}{Phys. Rev. C} \textbf{\bibinfo{volume}{65}},
  \bibinfo{pages}{054306} (\bibinfo{year}{2002}).

\bibitem[{\citenamefont{Frick and M{\"u}ther}(2003)}]{frick03}
\bibinfo{author}{\bibfnamefont{T.}~\bibnamefont{Frick}} \bibnamefont{and}
  \bibinfo{author}{\bibfnamefont{H.}~\bibnamefont{M{\"u}ther}},
  \bibinfo{journal}{Phys. Rev. C} \textbf{\bibinfo{volume}{68}},
  \bibinfo{pages}{034310} (\bibinfo{year}{2003}).

\bibitem[{\citenamefont{Frick et~al.}(2005)\citenamefont{Frick, M{\"u}ther,
  Rios, Polls, and Ramos}}]{frick05}
\bibinfo{author}{\bibfnamefont{T.}~\bibnamefont{Frick}},
  \bibinfo{author}{\bibfnamefont{H.}~\bibnamefont{M{\"u}ther}},
  \bibinfo{author}{\bibfnamefont{A.}~\bibnamefont{Rios}},
  \bibinfo{author}{\bibfnamefont{A.}~\bibnamefont{Polls}}, \bibnamefont{and}
  \bibinfo{author}{\bibfnamefont{A.}~\bibnamefont{Ramos}},
  \bibinfo{journal}{Phys. Rev. C} \textbf{\bibinfo{volume}{71}},
  \bibinfo{pages}{014313} (\bibinfo{year}{2005}).

\bibitem[{\citenamefont{Rios et~al.}(2006)\citenamefont{Rios, Polls, Ramos, and
  M{\"u}ther}}]{rios06}
\bibinfo{author}{\bibfnamefont{A.}~\bibnamefont{Rios}},
  \bibinfo{author}{\bibfnamefont{A.}~\bibnamefont{Polls}},
  \bibinfo{author}{\bibfnamefont{A.}~\bibnamefont{Ramos}}, \bibnamefont{and}
  \bibinfo{author}{\bibfnamefont{H.}~\bibnamefont{M{\"u}ther}},
  \bibinfo{journal}{Phys. Rev. C} \textbf{\bibinfo{volume}{74}},
  \bibinfo{pages}{054317} (\bibinfo{year}{2006}).

\bibitem[{\citenamefont{Yakovlev and Pethick}(2004)}]{yakovlev04}
\bibinfo{author}{\bibfnamefont{D.~G.} \bibnamefont{Yakovlev}} \bibnamefont{and}
  \bibinfo{author}{\bibfnamefont{C.~J.} \bibnamefont{Pethick}},
  \bibinfo{journal}{Annu. Rev. Astron. Astrophys.}
  \textbf{\bibinfo{volume}{42}}, \bibinfo{pages}{169} (\bibinfo{year}{2004}).

\bibitem[{\citenamefont{Oechslin and Janka}(2007)}]{janka07}
\bibinfo{author}{\bibfnamefont{R.}~\bibnamefont{Oechslin}} \bibnamefont{and}
  \bibinfo{author}{\bibfnamefont{H.-T.} \bibnamefont{Janka}},
  \bibinfo{journal}{Phys. Rev. Lett.} \textbf{\bibinfo{volume}{99}},
  \bibinfo{pages}{121102} (\bibinfo{year}{2007}).

\bibitem[{\citenamefont{Bo{\.z}ek}(1999{\natexlab{b}})}]{bozek99a}
\bibinfo{author}{\bibfnamefont{P.}~\bibnamefont{Bo{\.z}ek}},
  \bibinfo{journal}{Nucl. Phys. A} \textbf{\bibinfo{volume}{657}},
  \bibinfo{pages}{187} (\bibinfo{year}{1999}{\natexlab{b}}).

\bibitem[{\citenamefont{M{\"u}ther and Dickhoff}(2005)}]{dickhoff05}
\bibinfo{author}{\bibfnamefont{H.}~\bibnamefont{M{\"u}ther}} \bibnamefont{and}
  \bibinfo{author}{\bibfnamefont{W.~H.} \bibnamefont{Dickhoff}},
  \bibinfo{journal}{Phys. Rev. C} \textbf{\bibinfo{volume}{72}},
  \bibinfo{pages}{054313} (\bibinfo{year}{2005}).

\bibitem[{\citenamefont{Dieperink et~al.}(2003)\citenamefont{Dieperink, Dewulf,
  Van~Neck, Waroquier, and Rodin}}]{dewulf03a}
\bibinfo{author}{\bibfnamefont{A.~E.~L.} \bibnamefont{Dieperink}},
  \bibinfo{author}{\bibfnamefont{Y.}~\bibnamefont{Dewulf}},
  \bibinfo{author}{\bibfnamefont{D.}~\bibnamefont{Van~Neck}},
  \bibinfo{author}{\bibfnamefont{M.}~\bibnamefont{Waroquier}},
  \bibnamefont{and} \bibinfo{author}{\bibfnamefont{V.}~\bibnamefont{Rodin}},
  \bibinfo{journal}{Phys. Rev. C} \textbf{\bibinfo{volume}{68}},
  \bibinfo{pages}{064307} (\bibinfo{year}{2003}).

\bibitem[{\citenamefont{Machleidt et~al.}(1996)\citenamefont{Machleidt,
  Sammarruca, and Song}}]{cdbonn}
\bibinfo{author}{\bibfnamefont{R.}~\bibnamefont{Machleidt}},
  \bibinfo{author}{\bibfnamefont{F.}~\bibnamefont{Sammarruca}},
  \bibnamefont{and} \bibinfo{author}{\bibfnamefont{Y.}~\bibnamefont{Song}},
  \bibinfo{journal}{Phys. Rev. C} \textbf{\bibinfo{volume}{53}},
  \bibinfo{pages}{R1483} (\bibinfo{year}{1996}).

\bibitem[{\citenamefont{Wiringa et~al.}(1995)\citenamefont{Wiringa, Stoks, and
  Schiavilla}}]{av18}
\bibinfo{author}{\bibfnamefont{R.~B.} \bibnamefont{Wiringa}},
  \bibinfo{author}{\bibfnamefont{V.~G.~J.} \bibnamefont{Stoks}},
  \bibnamefont{and}
  \bibinfo{author}{\bibfnamefont{R.}~\bibnamefont{Schiavilla}},
  \bibinfo{journal}{Phys. Rev. C} \textbf{\bibinfo{volume}{51}},
  \bibinfo{pages}{38} (\bibinfo{year}{1995}).

\bibitem[{\citenamefont{Baldo and Maieron}(2008)}]{baldo08}
\bibinfo{author}{\bibfnamefont{M.}~\bibnamefont{Baldo}} \bibnamefont{and}
  \bibinfo{author}{\bibfnamefont{C.}~\bibnamefont{Maieron}},
  \bibinfo{journal}{Phys. Rev. C} \textbf{\bibinfo{volume}{77}},
  \bibinfo{pages}{015801} (\bibinfo{year}{2008}).

\bibitem[{\citenamefont{Ho}(2004)}]{ho04}
\bibinfo{author}{\bibfnamefont{T.-L.} \bibnamefont{Ho}},
  \bibinfo{journal}{Phys. Rev. Lett.} \textbf{\bibinfo{volume}{92}},
  \bibinfo{pages}{090402} (\bibinfo{year}{2004}).

\bibitem[{\citenamefont{Kohler}()}]{kohler08}
\bibinfo{author}{\bibfnamefont{H.~S.} \bibnamefont{Kohler}},
  \bibinfo{note}{arxiv:0705.0944 \& arxiv:0801.1123}.

\bibitem[{\citenamefont{Fetter and Walecka}(2003)}]{fetter}
\bibinfo{author}{\bibfnamefont{A.~L.} \bibnamefont{Fetter}} \bibnamefont{and}
  \bibinfo{author}{\bibfnamefont{J.~D.} \bibnamefont{Walecka}},
  \emph{\bibinfo{title}{Quantum Theory of Many-Particle Systems}}
  (\bibinfo{publisher}{Dover}, \bibinfo{address}{NY}, \bibinfo{year}{2003}).

\bibitem[{\citenamefont{Kadanoff and Baym}(1962)}]{kadanoff}
\bibinfo{author}{\bibfnamefont{L.~P.} \bibnamefont{Kadanoff}} \bibnamefont{and}
  \bibinfo{author}{\bibfnamefont{G.}~\bibnamefont{Baym}},
  \emph{\bibinfo{title}{Quantum Statistical Mechanics}}
  (\bibinfo{publisher}{Benjamin}, \bibinfo{address}{N.Y.},
  \bibinfo{year}{1962}).

\bibitem[{\citenamefont{Baym}(1962)}]{baym62}
\bibinfo{author}{\bibfnamefont{G.}~\bibnamefont{Baym}}, \bibinfo{journal}{Phys.
  Rev.} \textbf{\bibinfo{volume}{127}}, \bibinfo{pages}{1391}
  (\bibinfo{year}{1962}).

\bibitem[{\citenamefont{Frick}(2004)}]{frickphd}
\bibinfo{author}{\bibfnamefont{T.}~\bibnamefont{Frick}}, Ph.D. thesis,
  \bibinfo{school}{University of T{\"u}bingen} (\bibinfo{year}{2004}).

\bibitem[{\citenamefont{Rios}(2007)}]{riosphd}
\bibinfo{author}{\bibfnamefont{A.}~\bibnamefont{Rios}}, Ph.D. thesis,
  \bibinfo{school}{University of Barcelona} (\bibinfo{year}{2007}).

\bibitem[{\citenamefont{Thouless}(1960)}]{thouless60}
\bibinfo{author}{\bibfnamefont{D.~J.} \bibnamefont{Thouless}},
  \bibinfo{journal}{Ann. Phys.} \textbf{\bibinfo{volume}{10}},
  \bibinfo{pages}{553} (\bibinfo{year}{1960}).

\bibitem[{\citenamefont{Alm et~al.}(1996)\citenamefont{Alm, R{\"o}pke, Schnell,
  Kwong, and K{\"o}hler}}]{alm96}
\bibinfo{author}{\bibfnamefont{T.}~\bibnamefont{Alm}},
  \bibinfo{author}{\bibfnamefont{G.}~\bibnamefont{R{\"o}pke}},
  \bibinfo{author}{\bibfnamefont{A.}~\bibnamefont{Schnell}},
  \bibinfo{author}{\bibfnamefont{N.~H.} \bibnamefont{Kwong}}, \bibnamefont{and}
  \bibinfo{author}{\bibfnamefont{H.~S.} \bibnamefont{K{\"o}hler}},
  \bibinfo{journal}{Phys. Rev. C} \textbf{\bibinfo{volume}{53}},
  \bibinfo{pages}{2181} (\bibinfo{year}{1996}).

\bibitem[{\citenamefont{Galitskii and Migdal}(1958)}]{migdal58}
\bibinfo{author}{\bibfnamefont{V.~M.} \bibnamefont{Galitskii}}
  \bibnamefont{and} \bibinfo{author}{\bibfnamefont{A.~B.}
  \bibnamefont{Migdal}}, \bibinfo{journal}{JETP} \textbf{\bibinfo{volume}{7}},
  \bibinfo{pages}{96} (\bibinfo{year}{1958}).

\bibitem[{\citenamefont{Koltun}(1974)}]{koltun74}
\bibinfo{author}{\bibfnamefont{D.~S.} \bibnamefont{Koltun}},
  \bibinfo{journal}{Phys. Rev.} \textbf{\bibinfo{volume}{9}},
  \bibinfo{pages}{484} (\bibinfo{year}{1974}).

\bibitem[{\citenamefont{Luttinger and Ward}(1960)}]{luttinger60b}
\bibinfo{author}{\bibfnamefont{J.~M.} \bibnamefont{Luttinger}}
  \bibnamefont{and} \bibinfo{author}{\bibfnamefont{J.~C.} \bibnamefont{Ward}},
  \bibinfo{journal}{Phys. Rev.} \textbf{\bibinfo{volume}{118}},
  \bibinfo{pages}{1417} (\bibinfo{year}{1960}).

\bibitem[{\citenamefont{Carneiro and Pethick}(1975)}]{carneiro75}
\bibinfo{author}{\bibfnamefont{G.~M.} \bibnamefont{Carneiro}} \bibnamefont{and}
  \bibinfo{author}{\bibfnamefont{C.~J.} \bibnamefont{Pethick}},
  \bibinfo{journal}{Phys. Rev. B} \textbf{\bibinfo{volume}{11}},
  \bibinfo{pages}{1106} (\bibinfo{year}{1975}).

\bibitem[{\citenamefont{Nicotra}(2005)}]{nicotraphd}
\bibinfo{author}{\bibfnamefont{O.~E.} \bibnamefont{Nicotra}}, Ph.D. thesis,
  \bibinfo{school}{University of Catania} (\bibinfo{year}{2005}).

\bibitem[{\citenamefont{Baldo and Ferreira}(1999)}]{baldo99}
\bibinfo{author}{\bibfnamefont{M.}~\bibnamefont{Baldo}} \bibnamefont{and}
  \bibinfo{author}{\bibfnamefont{L.~S.} \bibnamefont{Ferreira}},
  \bibinfo{journal}{Phys. Rev. C} \textbf{\bibinfo{volume}{59}},
  \bibinfo{pages}{682} (\bibinfo{year}{1999}).

\bibitem[{\citenamefont{Rios et~al.}(2005)\citenamefont{Rios, Polls, Ramos, and
  Vida{\~n}a}}]{rios05}
\bibinfo{author}{\bibfnamefont{A.}~\bibnamefont{Rios}},
  \bibinfo{author}{\bibfnamefont{A.}~\bibnamefont{Polls}},
  \bibinfo{author}{\bibfnamefont{A.}~\bibnamefont{Ramos}}, \bibnamefont{and}
  \bibinfo{author}{\bibfnamefont{I.}~\bibnamefont{Vida{\~n}a}},
  \bibinfo{journal}{Phys. Rev. C} \textbf{\bibinfo{volume}{72}},
  \bibinfo{pages}{024316} (\bibinfo{year}{2005}).

\bibitem[{\citenamefont{Mukherjee and Pandharipande}(2007)}]{mukherjee07}
\bibinfo{author}{\bibfnamefont{A.}~\bibnamefont{Mukherjee}} \bibnamefont{and}
  \bibinfo{author}{\bibfnamefont{V.~R.} \bibnamefont{Pandharipande}},
  \bibinfo{journal}{Phys. Rev. C} \textbf{\bibinfo{volume}{75}},
  \bibinfo{pages}{035802} (\bibinfo{year}{2007}).

\bibitem[{\citenamefont{Huang}(1987)}]{huang87}
\bibinfo{author}{\bibfnamefont{K.}~\bibnamefont{Huang}},
  \emph{\bibinfo{title}{Statistical Mechanics}} (\bibinfo{publisher}{John Wiley
  and Sons}, \bibinfo{address}{N.Y.}, \bibinfo{year}{1987}),
  \bibinfo{edition}{2nd} ed.

\bibitem[{\citenamefont{Horowitz and Schwenk}(2006)}]{schwenk06}
\bibinfo{author}{\bibfnamefont{C.}~\bibnamefont{Horowitz}} \bibnamefont{and}
  \bibinfo{author}{\bibfnamefont{A.}~\bibnamefont{Schwenk}},
  \bibinfo{journal}{Phys. Lett. B} \textbf{\bibinfo{volume}{638}},
  \bibinfo{pages}{153} (\bibinfo{year}{2006}).

\bibitem[{\citenamefont{Tol{\'o}s et~al.}(2008)\citenamefont{Tol{\'o}s, Friman,
  and Schwenk}}]{tolos08}
\bibinfo{author}{\bibfnamefont{L.}~\bibnamefont{Tol{\'o}s}},
  \bibinfo{author}{\bibfnamefont{B.}~\bibnamefont{Friman}}, \bibnamefont{and}
  \bibinfo{author}{\bibfnamefont{A.}~\bibnamefont{Schwenk}},
  \bibinfo{journal}{Nucl. Phys. A} \textbf{\bibinfo{volume}{806}},
  \bibinfo{pages}{105} (\bibinfo{year}{2008}).

\bibitem[{\citenamefont{Luttinger}(1961)}]{luttinger61}
\bibinfo{author}{\bibfnamefont{J.~M.} \bibnamefont{Luttinger}},
  \bibinfo{journal}{Phys. Rev.} \textbf{\bibinfo{volume}{121}},
  \bibinfo{pages}{942} (\bibinfo{year}{1961}).

\bibitem[{\citenamefont{Zuo et~al.}(2006)\citenamefont{Zuo, Li, Lombardo, Lu,
  and Schulze}}]{zuo06}
\bibinfo{author}{\bibfnamefont{W.}~\bibnamefont{Zuo}},
  \bibinfo{author}{\bibfnamefont{Z.~H.} \bibnamefont{Li}},
  \bibinfo{author}{\bibfnamefont{U.}~\bibnamefont{Lombardo}},
  \bibinfo{author}{\bibfnamefont{G.~C.} \bibnamefont{Lu}}, \bibnamefont{and}
  \bibinfo{author}{\bibfnamefont{H.-J.} \bibnamefont{Schulze}},
  \bibinfo{journal}{Phys. Rev. C} \textbf{\bibinfo{volume}{73}},
  \bibinfo{pages}{035208} (\bibinfo{year}{2006}).

\bibitem[{\citenamefont{Bo{\.z}ek and Czerski}(2002)}]{bozek02a}
\bibinfo{author}{\bibfnamefont{P.}~\bibnamefont{Bo{\.z}ek}} \bibnamefont{and}
  \bibinfo{author}{\bibfnamefont{P.}~\bibnamefont{Czerski}},
  \bibinfo{journal}{Phys. Rev. C} \textbf{\bibinfo{volume}{66}},
  \bibinfo{pages}{027301} (\bibinfo{year}{2002}).

\bibitem[{\citenamefont{Negele and Orland}(1998)}]{negele}
\bibinfo{author}{\bibfnamefont{J.~W.} \bibnamefont{Negele}} \bibnamefont{and}
  \bibinfo{author}{\bibfnamefont{H.}~\bibnamefont{Orland}},
  \emph{\bibinfo{title}{Quantum Many-Particle Systems}}
  (\bibinfo{publisher}{Perseus}, \bibinfo{year}{1998}).

\bibitem[{\citenamefont{Ashcroft and Mermin}(1976)}]{ashcroft}
\bibinfo{author}{\bibfnamefont{N.~W.} \bibnamefont{Ashcroft}} \bibnamefont{and}
  \bibinfo{author}{\bibfnamefont{N.~D.} \bibnamefont{Mermin}},
  \emph{\bibinfo{title}{Solid State Physics}} (\bibinfo{publisher}{Holt,
  Rinehart and Winston}, \bibinfo{year}{1976}).

\bibitem[{\citenamefont{Grang\'e et~al.}(1987)\citenamefont{Grang\'e, Cugnon,
  and Lejeune}}]{grange87}
\bibinfo{author}{\bibfnamefont{P.}~\bibnamefont{Grang\'e}},
  \bibinfo{author}{\bibfnamefont{J.}~\bibnamefont{Cugnon}}, \bibnamefont{and}
  \bibinfo{author}{\bibfnamefont{A.}~\bibnamefont{Lejeune}},
  \bibinfo{journal}{Nucl. Phys. A} \textbf{\bibinfo{volume}{473}},
  \bibinfo{pages}{365} (\bibinfo{year}{1987}).

\end{thebibliography}

\end{document}